\documentclass[twocolumn,floatfix]{aastex631}
\usepackage{bm}
\usepackage{upgreek}
\usepackage{float}
\graphicspath{ {./Images/} }

\shorttitle{Early Stage of Protostellar Disk}
\shortauthors{\textsc{Noel et al.}}

\begin{document}

\title{Early Stages of Protostellar Disk Evolution: A Link to the Initial Cloud Core}

\correspondingauthor{Majd Noel, Shantanu Basu}
\email{mnoel25@uwo.ca, basu@cita.utoronto.ca}

\author[0009-0004-2453-3677]{Majd Noel}
\affiliation{Department of Physics and Astronomy,
University of Western Ontario,
London, Ontario N6A 3K7, Canada}
%\email{mnoel25@uwo.ca}
\affiliation{Canadian Institute for Theoretical Astrophysics, University of Toronto, 60 Saint George St., Toronto, ON M5S 3H8, Canada}
\author[0009-0001-3446-8440]{Rahul Khanna}
\affiliation{Department of Physics and Astronomy,
University of Western Ontario,
London, Ontario N6A 3K7, Canada}
\email{rkhann43@uwo.ca}
\author[0000-0003-0428-2140]{Shahram Abbassi}
\affiliation{Department of Physics and Astronomy,
University of Western Ontario,
London, Ontario N6A 3K7, Canada}
\email{sabbassi@uwo.ca}
\author[0000-0002-8697-9808]{Sami Dib}
\affiliation{Max Planck Institute for Astronomy, Königstuhl 17, D-69117, Heidelberg, Germany}
\email{sami.dib@gmail.com}
\author[0000-0003-0855-350X]{Shantanu Basu}
\affiliation{Department of Physics and Astronomy,
University of Western Ontario,
London, Ontario N6A 3K7, Canada}
\affiliation{Canadian Institute for Theoretical Astrophysics, University of Toronto, 60 Saint George St., Toronto, ON M5S 3H8, Canada}
%\email{basu@cita.utoronto.ca}

%\author
{
%Erfan Nourbakhsh\altaffilmark{1,2 $\star$},
%Sami Dib\altaffilmark{3,4 $\dagger$},
%Shahram Abbassi\altaffilmark{5,2 $\ddagger$},
%Habib G. Khosroshahi\altaffilmark{2 $\mathsection$},
}

% \affil{$^1$ Department of Physics, University of California Davis, One Shields Avenue, Davis, California 95616, USA}
% \affil{$^2$ School of Astronomy, Institute for Research in Fundamental Sciences (IPM), Tehran, 19395-5531, Iran}
% \affil{$^3$ Niels Bohr International Academy, Niels Bohr Institute, Blegdamsvej 17, DK-2100, Copenhagen, Denmark}
% \affil{$^4$ Centre for Star and Planet Formation, University of Copenhagen, {\O}ster Voldgade 5-7, DK-1350, Copenhagen, Denmark}
% \affil{$^5$ Department of Physics, School of Sciences, Ferdowsi University of Mashhad, Mashhad, 91775-1436, Iran}
% \affil{$^6$ Niels Bohr Institute, University of Copenhagen, Juliane Maries Vej 30, DK-2100 Copenhagen, Denmark}

% \email{$^\star$ EN: nourbakhsh@ucdavis.edu}
% \email{$^\dagger$ SD: sdib@nbi.dk}
% \email{$^\ddagger$ SA: abbassi@um.ac.ir}
% \email{$^\mathsection$ HGK: habib@ipm.ir}
% \email{$^\mathparagraph$ JJ: jeskj@nbi.dk}

\begin{abstract}
We study the structure and evolution of the very early protostellar disk (``protodisk'') just after protostar formation, where disk self-gravity dominates and the stellar contribution is dynamically minor. 
The disk redistributes angular momentum outward through outflows and gravitational torques, thereby helping to resolve the angular momentum problem of star formation.
We develop a self-similar model and carry out a parameter study that examines disk stability as a function of the key drivers of early evolution, notably the infall rate from the envelope and the strength of the gravitational torques.
The mass infall rate onto the disk is estimated to be that from the collapse of a Bonnor-Ebert sphere.
Our results indicate that protostellar disks that form from more unstable initial cores are more likely to be Toomre-unstable. 
We also find that the specific angular momentum of young protostellar disks lie in the range $10^{19}\text{--}10^{20}\,{\rm cm^2\,s^{-1}}$. 
We find distinct power-law profiles of physical quantities in the protodisk stage, including a rotation velocity profile that is shallower than the Keplerian profile that would be established at a later stage.
As a rough validity window, our assumptions are most secure during the first $\lesssim 2\times 10^{3}$\,yr after protostar formation and may plausibly extend to $\sim(0.5\text{--}1)\times 10^{4}$\,yr under weak magnetic braking and strong infall.
\end{abstract}

\keywords{Star formation (1569) --- Stellar winds (1636) --- Circumstellar disks (235) --- Hydrodynamics (1963) --- Gravitational collapse (662) --- Interstellar medium (847) --- Stellar accretion disks (1579)}

%\keywords{accretion, accretion disks --- ISM: clouds --- stars: formation --- stars: protostars}

\section{Introduction} \label{intro}

The earliest phases of star and planet formation are governed by the formation and evolution of a rotationally supported disk that mediates mass and angular momentum flow from an infalling envelope onto a newborn protostar. In the classical picture of inside–out collapse from a centrally condensed core, rotation and pressure set the conditions for disk formation and growth \citep{shu1977,TerebeyShuCassen1984}. Isothermal cores that resemble Bonnor-Ebert spheres \citep{bonnor1956mnras, ebert1955temperatur} located within observable filamentary column density structures are now widely regarded as plausible initial conditions for star formation in nearby molecular clouds \citep{andre2019araa}.

Direct observation of these embedded systems is challenging because the protostar and disk are enshrouded by a dense, optically thick envelope \citep{andre1993submillimeter}. Nevertheless, high-resolution millimeter interferometry has begun to resolve disks at the Class~0/I stages and to constrain their sizes, masses, and kinematics \citep{enoch2011disk,tobin20120,hara2013rotating,murillo2013keplerian,codella2014alma,lindberg2014alma, ohashi2013, ohashi2023early, aso2015alma, aso2019protostellar, maret2020searching}. Large-sample surveys indicate that Class~0 disks are typically compact (tens of au) rather than extended, and that large ($\gtrsim 60$~au) disks are the minority \citep{segura2018vla,maury2019characterizing,tobin2020vandam,ohashi2023early,hsieh2024alma}. These demographics, together with dust-based mass estimates, imply that by the Class~0/I epochs the central protostar often rivals or exceeds the disk in mass, even as inflow from the envelope continues.

Theoretical and numerical studies point to highly time-dependent accretion during the embedded phase, with elevated infall rates shortly after protostar formation and subsequent decline as the envelope is depleted \citep{masunaga2000radiation,schmeja2004protostellar,dib2010imf}. When the disk is sufficiently massive, self-gravity drives spiral arms and global asymmetries that transport angular momentum \citep{lin1987grav}, and can trigger fragmentation and burst-mode accretion \citep{vorobyov2005origin, vorobyov2006burst, vorobyov2010}. In parallel, magnetic processes regulate both disk formation and angular momentum loss. In the dense, poorly ionized midplane, non-ideal magnetohydrodynamic (MHD) effects (ohmic dissipation, ambipolar diffusion, and the Hall effect) alter magnetic braking, favor disk formation, and can depend on specific field–rotation geometries \citep{machida2011effect,tsukamoto2015effects,masson2016ad,wurster2020dominant,basu2024pd}. 

Jets and wide-angle winds provide an additional channel for angular momentum extraction. Observations increasingly support a tight accretion–ejection connection in Class~0/I systems, with outflows capable of removing angular momentum from the inner envelope and disk surfaces \citep{pudritz2019outflows,gaudel2020angular}. In combination, gravitational torques and magneto-centrifugal winds likely set the early angular momentum budget and regulate disk growth.

Despite this progress, analytical models that jointly (i) treat envelope-fed disks in the embedded stage, (ii) include self-gravity–driven transport, and (iii) allow for mass and angular momentum loss via outflows are still limited. Many classic treatments adopt isothermal collapse or prescribe turbulent viscosity without explicit gravitational instability (GI) physics \citep{TerebeyShuCassen1984,basu1997}. A physically motivated alternative is to parameterize transport by GI itself and to seek similarity solutions for non-isothermal, self-gravitating disks \citep{abbassi2006selfsimilar,abbassi2013viscous}. Those studies established a useful framework but employed idealized boundaries that were not explicitly tied to the thermodynamic state and mass loading of a pressure-confined parent core.

In this paper we build on that foundation and develop a self-similar, non-isothermal model for the \emph{very early} protostellar disk (“protodisk”) immediately after protostar formation. The model couples GI-regulated angular momentum transport to an outflow torque and links the disk’s outer boundary to a pressure-confined, Bonnor–Ebert–like core through an envelope mass-infall rate. We present time-dependent solutions for the surface density, temperature, specific angular momentum, and Toomre-$Q$, and we map the parameter space set by core sound speed, initial angular momentum content, and wind efficiency. The solutions quantify when and where the disk is gravitationally unstable, and they delineate the regimes that favor fragmentation, multiplicity, and bursty accretion.

Our scope is deliberately restricted to the nascent, self-gravity–dominated protodisk, when the stellar gravitational term and irradiation are dynamically minor. We therefore do not extrapolate these similarity solutions to typical Class~0/I systems, where surveys indicate that the star mass $M_\star$ exceeds the disk mass $M_{\rm disk}$ and magnetic/irradiative effects are increasingly important \citep{maury2019characterizing,tobin2020vandam}. We adopt our model for the first ${\lesssim}\,2\times10^{3}$~yr after the protostar formation, and could be evaluated potentially till $\sim(0.5$–$1)\times10^{4}$~yr under weak magnetic braking and strong infall; beyond this, explicit inclusion of a point-mass term and irradiation is warranted \citep{shu1977,TerebeyShuCassen1984,tomida2015rmhd,tsukamoto2015effects,masson2016ad}.

In Section~\ref{model}, we present the governing equations and theoretical formalism. Section~\ref{accrate} describes the accretion boundary condition imposed by the envelope. In Section~\ref{fiducial}, we discuss a representative solution and its time evolution. A broad parameter study is presented in Section~\ref{paramstudy}, followed by an analysis of angular momentum in Section~\ref{angmom}. In Section ~\ref{Simul}, we compare with early stage protostellar simulations, and then we conclude with a discussion of implications in Section~\ref{conclusion}.

\section{Model} \label{model}

Due to the substantial angular momentum of the parental cloud core, the formation of a centrifugally-supported disk is an inevitable consequence around the newborn protostar. The equations governing the physical behavior of such accretion disks are inherently complex and challenging to solve in their full form. Therefore, we employ a self-similar approach to simplify these equations, facilitating a more tractable analysis.

Our model assumes a geometrically thin ($H/r < 1$) and axisymmetric ($\partial / \partial \phi = 0$) accretion disk around a forming protostar, where the gravitational force is dominated by the disk's self-gravity, valid under the condition that the protostar mass $M_{\rm protostar} \ll M_{\rm disk}$. We intend this limit to describe the immediate post–protostar-formation, the protodisk stage, and accordingly we do not extrapolate these solutions to typical Class~0/I systems.
Winds and outflows are ubiquitously observed in embedded sources \citep{pudritz2019outflows,gaudel2020angular}, and are a common feature in numerical simulations, whether performed in local shearing boxes \citep{fromang2013local} or in global models of core collapse \citep{machida2019first, machida2024wind}.
%consistent with observations from numerical simulations of turbulent accretion disks \citep{fromang2013local} \textbf{and with the observed accretion–ejection connection in embedded sources \citep{pudritz2019outflows,gaudel2020angmom}}.
%Recent studies have emphasized the critical role of protostellar outflows in disk evolution. For instance, \citet{lebreuilly2024influence} found that outflows significantly impact both star and disk formation by injecting kinetic energy into the surrounding medium. Additionally, \citet{machida2019first} demonstrated through simulations that disk-driven outflows can effectively transport angular momentum, promoting disk growth and stability. 
These results motivate the simple wind parameterization adopted here and its inclusion alongside gravitational torques in the early envelope-fed stage.

The structural and temporal evolution of the disk is described by the vertically integrated equations of continuity, radial momentum, conservation of angular momentum, and mass loss rate:

\begin{equation}
\frac{\partial \sigma}{\partial t}
+\frac{1}{r}\frac{\partial}{\partial r}(r\sigma v_{r})
+2\dot{\sigma}_{\rm w}=0,
\label{eq1}
\end{equation}

\begin{equation}
\frac{\partial v_{r}}{\partial t}+v_{r} \frac{\partial v_{r}}{\partial r}
-\frac{j^2}{r^3}=-\frac{c_{\rm s}^2}{\sigma}\frac{\partial \sigma}{\partial r}
-\frac{GM_{r}}{r^2},
\label{eq2}
\end{equation}

\begin{equation}
\frac{\partial j}{\partial t}
+v_{r}\frac{\partial j}{\partial r}
=\frac{1}{r\sigma}\frac{\partial}{\partial r} \Big[r^3\sigma\nu\frac{\partial}{\partial r}(\frac{j}{r^2})\Big]
-\frac{2j}{\sigma}(l^2 \dot{\sigma}_{\rm w}),
\label{eq3}
\end{equation}

\begin{equation}
\frac{\partial \dot{M}_{\rm w}}{\partial r}=4 \pi r \dot{\sigma}_{\rm w},
\label{eq4}
\end{equation}
where $v_{r}$, $\sigma$, $j$, ${\nu}$, $c_{\rm s}$, $M_r$, and $\dot{M}_{\rm w}$ represent the radial velocity, surface density, specific angular momentum, effective viscosity, sound speed, mass enclosed within a radius $r$, and mass-loss rate due to disk outflows, respectively.
The parameter ${l}$ determines the strength of the extraction for the angular momentum and 
\begin{equation}
\dot{\sigma}_{\rm w}=\rho v_{z}^{+},
\label{eq5}
\end{equation}
is the mass loss rate per unit area from each disk face, where $v_{z}^{+}$ denotes the vertical wind velocity at its base, and $\rho$ is the mass density. The surface density is expressed by 
\begin{equation}
\sigma = 2\rho H,
\label{surface density}
\end{equation}
as in \cite{abbassi2013viscous}, where
\begin{equation}
H=\frac{c_{\rm s}^2}{2\pi G \sigma}
\label{eq6}
\end{equation}
is the scale height of the disk.
The pressure $p$ follows a polytropic relation,
\begin{equation}
p=K\rho^{\gamma},
\label{eq7}
\end{equation}
where $K$ is a constant for a given polytropic index $\gamma$.

Note that the gravitational field of a thin disk is not exactly equal to the value for a spherical distribution that is used here, but the difference is of order unity and the scaling with radius $r$ is expected to be the same. See Appendix \ref{app2} for further discussion.
To nondimensionalize the equations, we introduce a similarity variable \citep{yahil1983self, mineshige1997self}:
\begin{equation}
x = K^{-\frac{1}{2}}G^{\frac{\gamma-1}{2}} r ~ t^{\gamma-2}.
\label{eq8}
\end{equation}
This transformation reduces the partial differential equations to ordinary differential equations (ODEs) in terms of the similarity variable $x$. As discussed in \citet{abbassi2013viscous}, these solutions are valid under the assumptions of disk self-gravity dominance and the slow accretion limit. In line with our stated scope, gravity in Eq.~(\ref{eq2}) arises from the enclosed \emph{disk} mass $M_r$ only; the stellar point-mass term and irradiation are omitted because we restrict attention to the protodisk stage.

In the early stages, circumstellar disks are susceptible to GI, which can effectively transport angular momentum outward through gravitational torques \citep{larson1984gravitational, nomura2000angular, gammie2001nonlinear}. The effective viscosity $\nu_{\rm GI}$ due to GI is expressed as
\begin{equation}
\nu_{\rm GI}=\eta \frac{G \sigma^2 r^6}{j^3}
\label{eq9}
\end{equation}
\citep{lin1987grav},
where $\eta$ is a dimensionless parameter representing the efficiency of angular momentum transport. For an isothermal self-similar disk, \citet{nomura2000angular} reformulated this as $\nu^{\prime}_{\rm GI}=\eta^{\prime}\Sigma^2 x^6/J^3$, highlighting its dominance over other viscosity mechanisms in gravitationally unstable disks, where $\nu'_{\rm GI}$ and $\eta'$ are dimensionless variables in self-similar space. The variables $\Sigma$ and $J$ are dimensionless similarity variables corresponding to the surface density $\sigma$ and specific angular momentum $j$.

After appropriate mathematical manipulations, the primary ODE describing disk evolution becomes
\begin{eqnarray}
\frac{dV_{r}}{dx} & = & \frac{\frac{A}{{\nu_{\rm GI}^{\prime}}} + BV_{r} + Cx + 6 \left (\frac{V^{2}_{r}}{x} \right)} {6V_{r}-(3\gamma+2)x-4x\Gamma},
\label{eq:mainalpha}
\end{eqnarray}
with the coefficients 
\begin{eqnarray}
A & = & \frac{1}{x^2} [V_{r}-(2-\gamma)x]^4[V_{r}+(2l^2-1)x\Gamma], \\
B & = & x \frac{d\Gamma}{dx}-3\Gamma+12\gamma-18,\\ 
C & = & (\gamma-2)x\frac{d \Gamma}{dx}-2\Gamma^2+(5\gamma -2)\Gamma-10\gamma+12.
\label{eq10}
\end{eqnarray}
The additional ODEs for $\Sigma$ and $\dot{\mathcal{M}}_{\rm w}$ follow similarly:

\begin{equation}
\frac{d\Sigma}{dx} = -\frac{\Sigma}{x} - \frac{\Sigma \left ( \frac{dV_r}{dx} +\Gamma - 2\gamma + 2  \right)}{V_r + (\gamma-2)x}
\label{eq:main_density}\, ,
\end{equation}

%\label{eq10}
%\end{eqnarray}
%The additional ODEs for $\Sigma$ and $\dot{\mathcal{M}}_{\rm w}$ follow similarly:
%\begin{equation}
%\frac{d\Sigma}{dx} = -\frac{\Sigma}{x} - \frac{\Sigma( \frac{dV_r}{dx} +\gamma - 2) - \Sigma(3\gamma-\Gamma-4)}{V_r + (\gamma-2)x}
%\label{eq:main_density}\, ,
%\end{equation}

\begin{equation}
\frac{d\dot{\mathcal{M}}_{\rm w}}{dx} = \Sigma\,  x\, \Gamma.
\label{eq:main_mu_capital}
\end{equation}
These are derived in \citet{abbassi2013viscous}. The variable $\dot{\mathcal{M}_{\rm w}}$ is the dimensionless similarity variable of the mass loss rate $\dot{M}_{\rm w}$, $V_{r}$ is the dimensionless similarity parameter describing the radial velocity $v_{r}$, and $\Gamma$ is a parameter that encapsulates the intensity of the disk winds (see Appendix \ref{app1} for self-similar scaling). We numerically integrate these ODEs with respect to the similarity variable $x$ to obtain the physical variables of interest.
The boundary conditions adopted are $v_{r}=0$ at the center ($x=0$) and mass infall rate $\dot{M}=\dot{M}_{\rm infall}$ at the outer boundary ($x=1$), where $\dot{M}_{\rm infall}$ is a constant envelope infall rate during the early disk evolution (see Section \ref{accrate}). This setup allows us to explore the impact of envelope feeding on the disk structure and stability. For the outflow mass loss profile, we assume a power-law form: $\Lambda =\Lambda_{0} x^{s}$ (see equation \ref{a19}) where $s$ is the scaling exponent. In the isothermal limit, $\Lambda$ and $\Gamma$ converge, aligning with the wind parameters described in \citet{shadmehri2009influence}. Consistent with \citet{mineshige1997self}, we impose a constraint on $\gamma$ to ensure negative and bounded radial velocities as $x \rightarrow \infty$, adopting $\gamma=1.1$ as our fiducial value, which lies between the isothermal ($\gamma=1$) and adiabatic ($\gamma=1.4$) cases. This choice is also consistent with a cool, GI-regulated protodisk in our fiducial runs.

\section{Parental Cloud Core and Accretion Rates}\label{accrate}

Near-infrared observations \citep[e.g.,][]{alves2001internal,schnee2005density} have reported that molecular cloud cores are centrally condensed and exhibit structures that resemble Bonnor-Ebert (BE) spheres. Motivated by this, as a realistic model for the parental cloud core that begins to collapse, we consider the density profile of a critical BE sphere. %with the dimensionless radius $\xi = 6.45$. 
The initial temperature of the isothermal core is $T = 10~K$, and the central density is $\rho_{\rm 0} = 8 \times 10^{-19}~{\rm g~cm^{-3}}$, yielding a cloud radius of $R_{\rm cl} = 0.03~{\rm pc}$. Although such dense cores are considered thermally supported and gravitationally stable, they can be driven to collapse by increased external pressure or enhanced mass loading, which is consistent with recent ALMA observations of marginally stable star-forming cores in clustered environments \citep{chen2024dynamics}. In our framework, the BE core sets the envelope mass supply to the disk during the protodisk evolutionary period modeled here.

To initiate collapse in our model, we scale the entire density profile by factors $f = 1.2$, $1.68$, and $2.8$. These correspond to cloud core masses of ${M}_{\rm cl} = 1.26$, $2.09$, and $4.5~{\rm M}_\odot$, respectively. The non-enhanced BE core has a mass of approximately $0.625~{\rm M}_\odot$. As shown by \cite{matsumoto2003fragmentation}, the density enhancement factor $f$ is related to the initial ratio of thermal to gravitational energy, $\alpha_0$, via $\alpha_0 = 0.84 / f$. Thus, our $f$-values correspond to $\alpha_0 = 0.7$, $0.5$, and $0.3$. These three cases span from marginally stable ($\alpha_0\!=\!0.7$) to strongly unstable ($\alpha_0\!=\!0.3$) parental cores, thereby bracketing the mass-loading conditions relevant to the protodisk phase.

We define three initial conditions, $\rho_0$, $t_{\rm ff}$ and $\dot{M}_{\rm infall}$ that allow us to investigate how different stability states in the parent core influence disk evolution. The free-fall time is given by
\begin{equation}
t_{\rm ff} = \left( \frac{3 \pi}{32 G \rho_{\rm 0}} \right)^{1/2}.
\end{equation}
For $\rho_0=8\times10^{-19}\,{\rm g\,cm^{-3}}$, this yields $t_{\rm ff}\!\approx\!7.4\times10^{4}\,{\rm yr}$, which we take as the characteristic timescale that sets the envelope mass supply during the first few kyr of disk evolution.

From this, we estimate the mass infall rate as
\begin{equation}
\dot{M}_{\rm infall} = \frac{M_{\rm cl}}{t_{\rm ff}} = \alpha_0^{-3/2} \times 10^{-5}~{\rm M}_\odot~{\rm yr}^{-1}
\label{eq11}
\end{equation}
\citep[see also][]{machida2011recurrent}.
This expression shows how the infall rate scales with the degree of GI in the initial core, encapsulated by $\alpha_0$: lower $\alpha_0$ (more unstable) leads to a higher $\dot{M}_{\rm infall}$. Using Eq.~(\ref{eq11}), the three cases above give $\dot{M}_{\rm infall}\!\approx\!1.7\times10^{-5}$, $2.8\times10^{-5}$, and $6.1\times10^{-5}\,{\rm M_\odot\,yr^{-1}}$, respectively. In our model, this dependence provides the key dynamical link between the thermodynamic state of the parental core and the mass loading of the forming disk; we hold $\dot{M}_{\rm infall}$ constant over the $\sim$kyr interval considered to isolate the protodisk dynamics, leaving time-variable infall to future work.

\section{A Fiducial Case} \label{fiducial}

\begin{figure}
\begin{center}
\includegraphics[width=\columnwidth,height=0.5\textwidth]{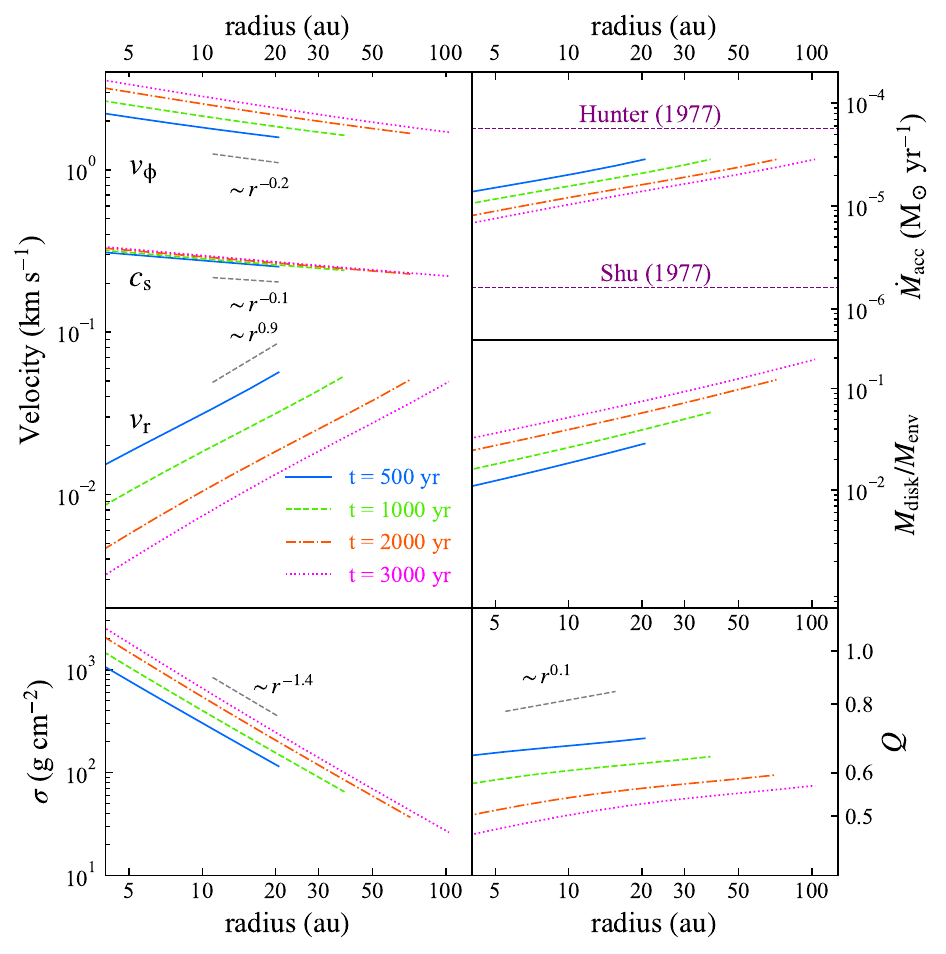}
\end{center}
\caption{The profiles of the physical variables at three epochs for $\alpha_0 = 0.5$, $\eta^{\prime}=10^{-2}$, $\gamma=1.1$ and a typical outflow emanating from the disk set at $(\Lambda_0,l,s)=(0.1, 1.5, 0.7)$. The panels show surface density, radial and tangential velocities, sound speed, and Toomre $Q$ profiles, illustrating both the radial and temporal evolution of the system.}
\label{timeevol}
\end{figure}

Figure~\ref{timeevol} presents the results of our fiducial model, with the parameters $\alpha_0 = 0.5$, $\eta^{\prime}=10^{-2}$, $\gamma=1.1$,  $\Lambda_0 = 0.1$, $ l = 1.5,$ and $s= 0.7$, illustrating the temporal evolution of disk structure at four time slices (0.5, 1, 2 and 3~kyr). For context, power-law fits are overlaid on several physical quantities to guide the eye and facilitate the interpretation. In our model, we consider the disk to be starting from $r > 5$ au, and the protostar to be $r < 5$ au, which satisfies our assumption that $M_* \ll M_{\rm disk}$ for evolutionary timescales under review. As time progresses, the surface density in the inner region increases, while the outer region—continuously fed by infalling envelope material—shows a declining surface-density profile as the disk spreads viscously. This behavior reflects outward angular-momentum redistribution, which drives radial disk growth over kyr timescales. Adopting $K = 80 \, \rm cm^{2.3}g^{-0.1}s^{-2}$, the outer radius in this fiducial solution reaches $\sim 40$~au at 1~kyr and $\sim 70$~au at 2~kyr; these values are consistent with very young Class 0 disk measurements reported in recent ALMA observations \citep{maureira2024faust}. These values are shown for orientation only, since our solutions target the protodisk limit $M_{\rm disk}\!\gg\!M_\star$.

The gradual increase of surface density at each radius signifies a steady mass build-up in the disk+protostar system. For instance, the surface-density profile at $t = 10^3~{\rm yr}$ follows a power law:
\begin{equation}
\sigma \simeq 7.78 \times 10^{3}~{\rm g~cm^{-2}}~\left( \frac{r}{\rm 1~au} \right)^{-1.4}
\label{powerlaw}
\end{equation}
%in good agreement with the results of \cite{ohtani2013growth}. 
(see also Appendix \ref{app2}).
Such normalizations and indices are observationally relevant, as they offer simplified parameterizations that can inform radiative-transfer modeling of very young disks. This agreement supports the internal consistency of our gravitational-torque–regulated model in the protodisk regime.

The tangential velocity and sound speed both increase over time in the inner regions, whereas radial velocity declines. The small ratio $|v_r|/v_\phi\!\ll\!1$ across radii confirms the slow-accretion approximation used in the similarity reduction, and the profiles remain geometrically thin ($H/r\!\ll\!1$). These features are consistent with self-similar viscous disk solutions \citep{ohtani2013growth}. 

To evaluate local gravitational stability, we adopt Toomre's $Q$ criterion \citep{toomre1964gravitational}:
\begin{equation}
Q=\frac{c_{\rm s} \kappa}{\pi G\sigma},
\label{toomre}
\end{equation}
where
\begin{equation}
\kappa = \Omega \left( 4 + 2 \frac{d \log \Omega}{d \log r} \right)^{1/2}
\label{epicyclic}
\end{equation}
is the epicyclic frequency and $\Omega= v_{\phi}/r$ is the angular velocity. Values of $Q > 1$ indicate local stability against axisymmetric perturbations, whereas $Q < 1$ signals GI. In our fiducial case, the $Q$ parameter generally decreases over time, primarily due to increasing surface density. Although the outer $\sigma$ decreases slightly (which by itself would raise $Q$), concurrent declines in $c_{\rm s}$ and $\kappa$ reduce the \emph{numerator} of $Q$ and drive a net decrease. A GI-active annulus with $Q\!\lesssim\!1$ therefore develops and broadens with time, providing the torque that redistributes angular momentum.

\begin{figure}
\begin{center}
\includegraphics[width=\columnwidth]{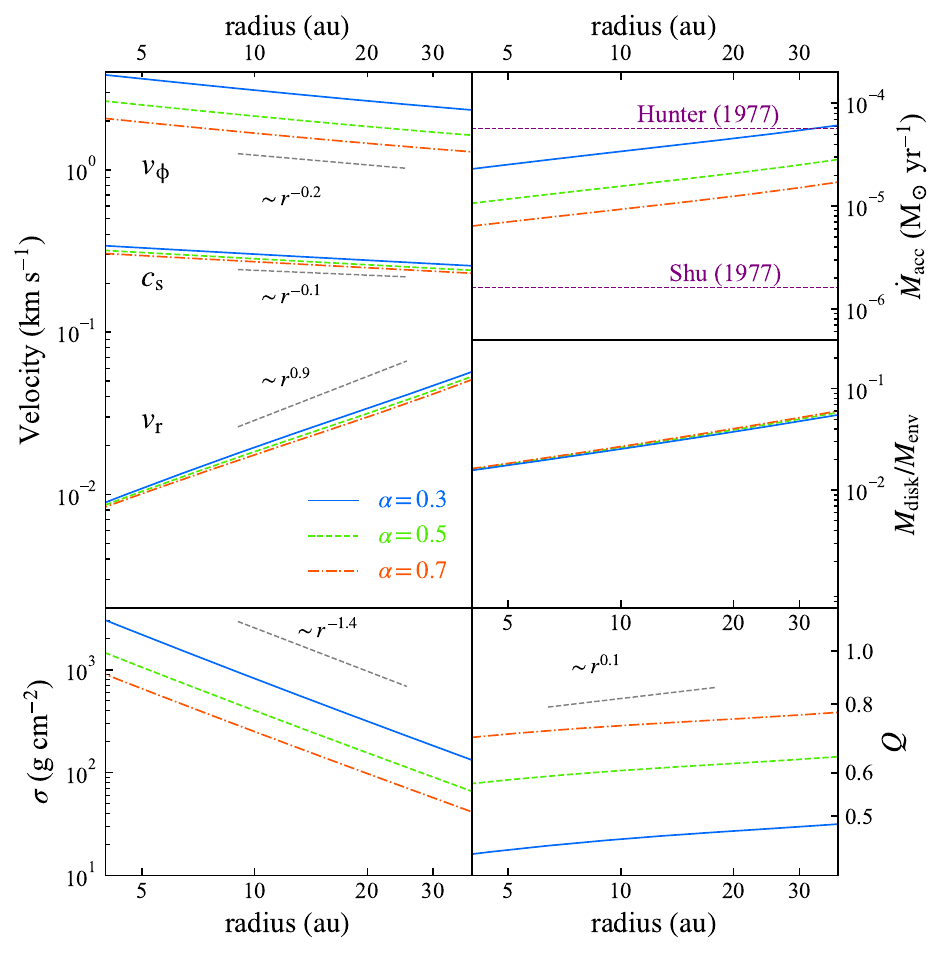}
\end{center}
\caption{The profiles of the physical variables at $t=10^3~{\rm yr}$ for $\alpha_0 = 0.7,0.5,0.3$ with $\eta=10^{-2}$, $\gamma=1.1$ and a typical outflow emanating from the disk set at $(\Lambda_0,l,s)=(0.1, 1.5, 0.7)$.}
\label{alpha}
\end{figure}

\begin{figure}
\begin{center}
\includegraphics[width=\columnwidth]{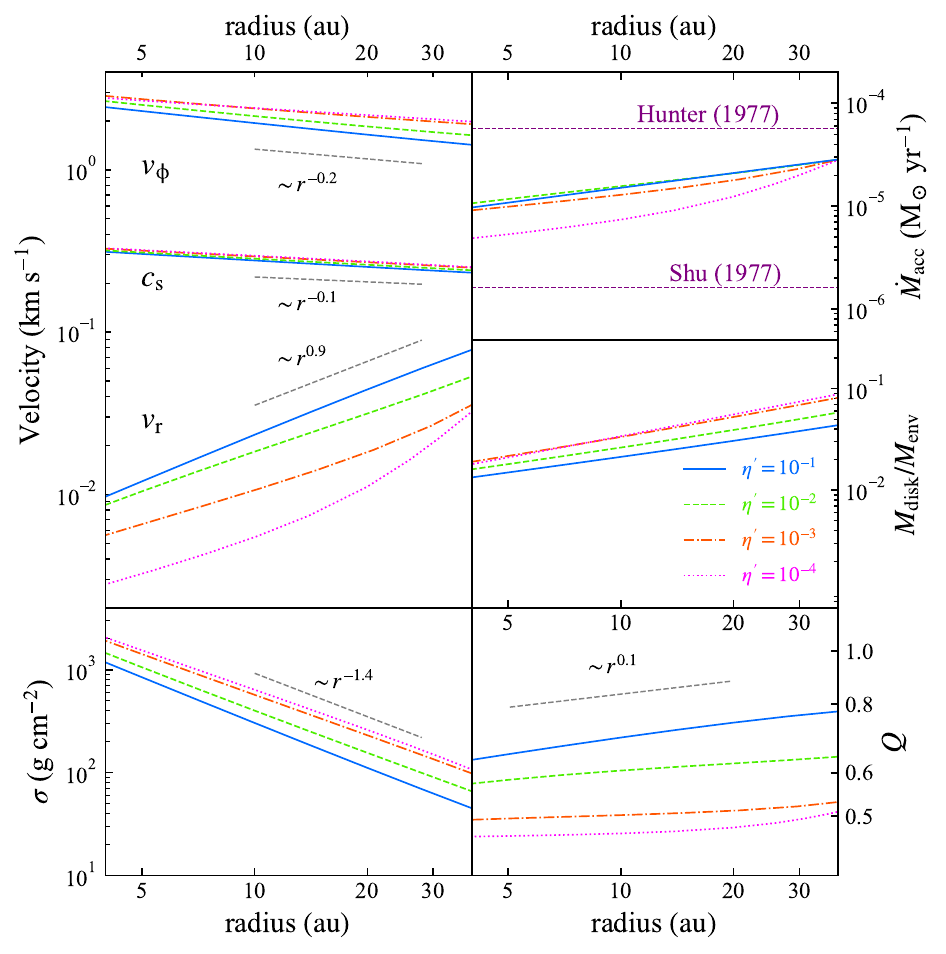}
\end{center}
\caption{Same as Figure \ref{alpha} but with fixed $\alpha_0 = 0.5$ and $\eta^{\prime}=10^{-1},10^{-2},10^{-3},10^{-4}$.}
\label{eta-prime}
\end{figure}

\begin{figure}
\begin{center}
\includegraphics[width=\columnwidth]{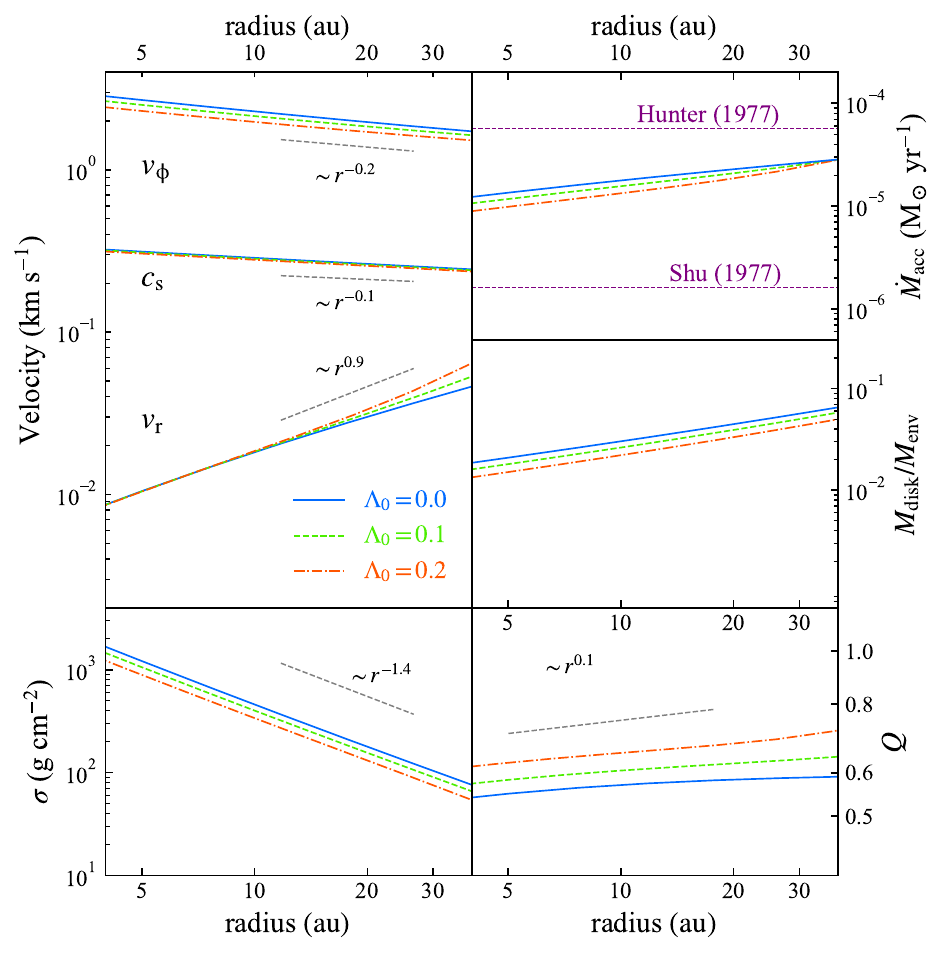}
\end{center}
\caption{Same as Figure \ref{alpha} but with fixed $\alpha_0 = 0.5$ and different strength attributed to outflows, i.e., $\Lambda_0=0,0.1,0.2$.}
\label{lambda}
\end{figure}

\begin{figure}
\begin{center}
\includegraphics[width=\columnwidth]{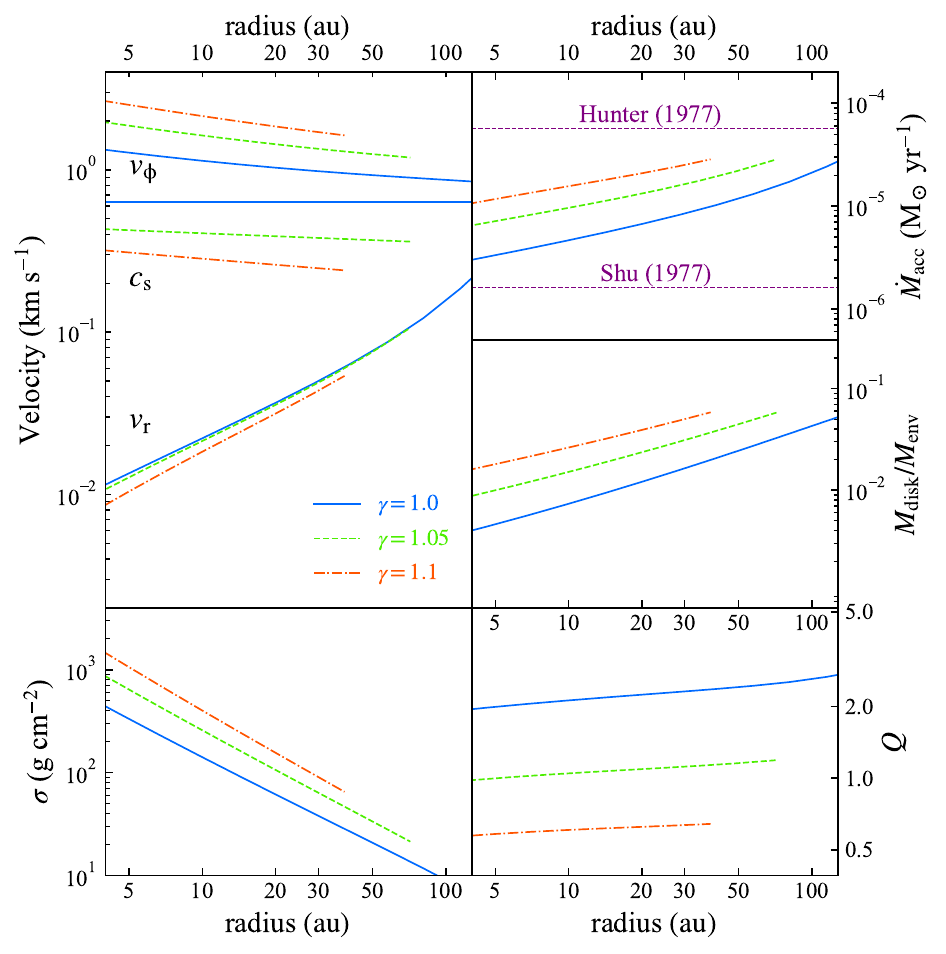}
\end{center}
\caption{Same as Figure \ref{alpha} but with fixed $\alpha_0 = 0.5$ and $\gamma=1.0,1.05,1.1$.}
\label{gamma}
\end{figure}

The envelope infall rate governs the mass growth of the disk. Under the assumption of a constant infall rate, the mass of the disk rises with time, because the accretion of the disk towards the protostar (will be discussed further in Section \ref{Simul}) is less than the accretion from the envelope towards the disk. The ratio ${M}_{\rm disk}/{M}_{\rm env}$ (middle-right panel of Figure~\ref{timeevol}) illustrates how the disk continuously accretes material from the envelope over the first few kyr of evolution. This behavior is consistent with the results of \cite{ohtani2013growth} and \cite{machida2013evolution}, both of which support efficient disk mass growth during early accretion phases. The rapid depletion of envelope mass reported in \cite{machida2013evolution} further supports our model’s implication that the main accretion phase is relatively short-lived in the protodisk stage.

Finally, it is insightful to compare our mass accretion rates with those derived from established collapse models. In general, $\dot{M}$ scales as $c_{\rm s}^3/G$, with the proportionality constant depending on the collapse mechanism: $\dot{M} \approx 0.975\, c_{\rm s}^3/G$ for inside-out collapse \citep{shu1977} and $\dot{M} \approx 36\, c_{\rm s}^3/G$ for early collapse stages in numerical simulations \citep{hunter1977collapse}. For $T = 10~\rm K$ ($c_{\rm s} = 0.19~\rm km~s^{-1}$), these yield $\dot{M} \approx 1.6 \times 10^{-6}~M_{\odot}~\rm yr^{-1}$ and $\dot{M} \approx 5.7 \times 10^{-5}~M_{\odot}~\rm yr^{-1}$, respectively—shown in Figures~\ref{timeevol}–\ref{gamma} for comparison. 
% We use these benchmarks for orientation only; our similarity solutions are intended for the protodisk limit and are not calibrated to typical Class~0/I demographics.

\section{Parameter Study}\label{paramstudy}

In this section, we explore the influence of key parameters in our model while fixing others to their fiducial values. We assess the sensitivity of disk structure and dynamics to four primary quantities: $\alpha_{0}$, $\eta^{\prime}$, $\Lambda_{0}$, and $\gamma$—presented in Figures~\ref{alpha}–\ref{gamma} for $t = 1~{\rm kyr}$. An essential aim is to investigate how the physical state of the parental molecular cloud core imprints itself on the emerging disk. This is mediated by $\alpha_{0}$, which encodes the ratio of thermal to gravitational energy in the core and sets the disk’s outer boundary condition (see Section~\ref{model}). Smaller $\alpha_{0}$ values correspond to more massive, gravitationally unstable cores, and are the main driver of variations in $\dot{M}_{\rm infall}$ in our study \citep{machida2011recurrent}. Recent observational surveys have also emphasized that Class 0 systems with higher envelope mass tend to exhibit more massive and potentially unstable disks \citep{tobin2020vandam, maury2019characterizing}. 
Figure~\ref{alpha} illustrates how varying $\alpha_{0}$ shapes radial profiles of disk quantities. The top-left panel reveals that lower $\alpha_{0}$ leads to enhanced tangential velocity and sound speed. At the outer disk edge, the sound speed approaches the isothermal limit ($\sim 0.2~{\rm km~s^{-1}}$), suggesting that gas becomes increasingly isothermal as it transitions from the disk to the envelope. This transition reflects the imposed boundary conditions, where the envelope was modeled with isothermal thermodynamics (see Section~\ref{accrate}). The radial velocity exhibits only weak sensitivity to $\alpha_{0}$. Higher inner disk sound speed indicates elevated temperatures toward the protostar, while the velocity profile confirms faster rotation in inner layers, consistent with angular momentum transport via GI and outflows. The right-top panel shows that lower $\alpha_{0}$ yields higher accretion rates and slightly elevated surface density (left-bottom panel). These trends follow directly from Eq.~(\ref{eq11}): smaller $\alpha_0$ implies a larger $\dot{M}_{\rm infall}$ and hence a more rapidly mass-loaded disk.

The right-bottom panel shows a clear trend of decreasing Toomre $Q$ with decreasing $\alpha_{0}$, as expected from equation~(\ref{eq11}). More unstable parental cores drive higher mass infall rates, which increase surface density and reduce $Q$, promoting GI \citep{lin1987grav}. Consequently, $Q \lesssim 1$ becomes more easily satisfied in these models. The ratio $M_{\rm disk}/M_{\rm env}$ hardly changes due to changing $\alpha_0$, because an increase (decrease) in the mass supplied to the disk is accompanied by an increase (decrease) in mass transport in the disk due to GI.
% The ratio $M_{\rm disk}/M_{\rm env}$ (slightly) increases with increasing $\alpha_{0}$. 
These trends are consistent with recent 3D simulations, which show that low thermal support during core collapse leads to the formation of more massive, gravitationally unstable disks \citep{tsukamoto2022fragmentation}, and that such disks can sustain GI-driven structure and fragmentation under realistic cooling conditions \citep{xu2024global}. In our similarity framework, this appears as a broadening GI-active annulus ($Q\!\lesssim\!1$) that supplies the torque for outward angular-momentum transport.

\begin{figure}
\begin{center}
\includegraphics[width=\columnwidth]{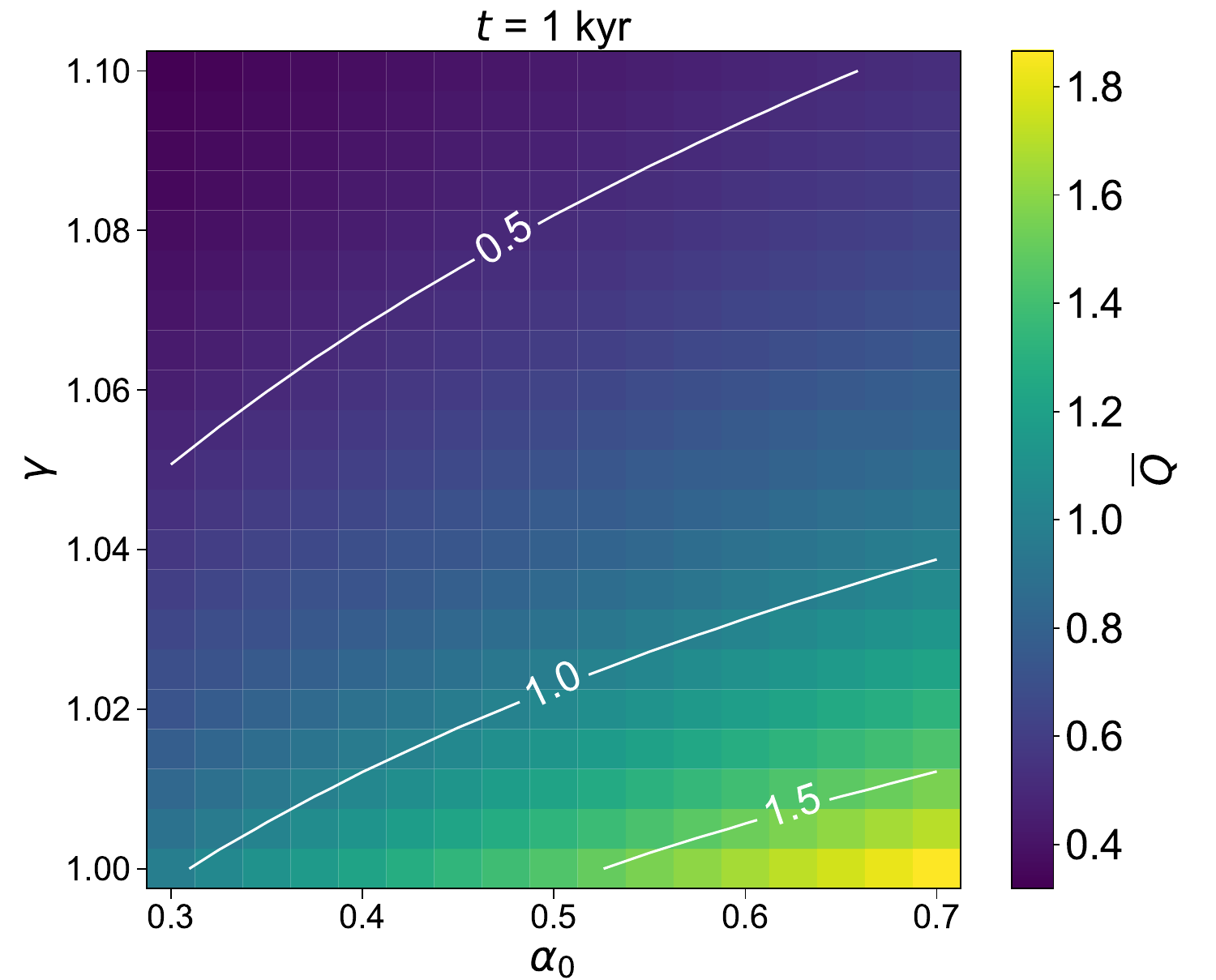}
\end{center}
\caption{Grid of models with two main parameters, $\alpha_0$ and $\gamma$ for a snapshot of $t = 1$ kyr. The white contours represent lines of constant $\bar{Q}$, which is the radially averaged $Q$.}
\label{grid}
\end{figure}

% \begin{figure}
% \begin{center}
% \includegraphics[width=\columnwidth]{f9.eps}
% \end{center}
% \caption{The time evolution of the fraction $M_{\rm disk}/M_{\rm env}$ plotted against $M_{\rm env}$. Each curve is labeled by the corresponding $\alpha_0$ value. Observational data from \citet{jorgensen2009prosac} are overlaid and split into Class 0 and I sources}
% \label{menv}
% \end{figure}

It is informative to analyze the radial distribution of Toomre's $Q$. Since $Q \propto c_s \kappa / \sigma$, and the surface density $\sigma$ varies more steeply with radius than the sound speed or epicyclic frequency, $Q$ typically peaks near the outer disk where $\sigma$ is lowest. Such behavior is consistent with other self-similar disk models \citep{shadmehri2009influence, abbassi2013viscous}, and is qualitatively similar to the isothermal collapse scenario studied by \cite{nomura2000angular}. However, it differs from \cite{ohtani2013growth}, who found higher $Q$ in the inner disk—likely due to their stiffer polytropic index ($\gamma = 1.4$) and inclusion of a central accreting protostar. In such scenarios, protostellar radiation can suppress inner disk instabilities. In contrast, our models—e.g., those shown in Figure~\ref{alpha}—maintain $Q < 1$ throughout much of the disk, preserving GI and gravitational torque-driven viscosity. Recent ALMA imaging has revealed spiral structures in Class 0 disks that also point to sustained GI across large radii \citep{segura2020alma}. 

% :):D

Figure~\ref{eta-prime} shows the impact of the effective GI viscosity parameter $\eta'$ on the disk. Increasing $\eta^{\prime}$—our proxy for GI efficiency—enhances angular momentum transport. This causes slight decreases in tangential velocity, sound speed, and surface density, while $Q$ increases across the disk. Higher $\eta^{\prime}$ reduces the disk’s capacity to retain material, thereby decreasing the accretion rate and increasing the outward radial velocity at the disk edge. Physically, larger $\eta'$ corresponds to stronger spiral torques that more efficiently drain mass inward and angular momentum outward.
% removed flattening the profile

Figure~\ref{lambda} examines the effect of outflows via the wind mass-loading parameter $\Lambda_0$. Higher $\Lambda_0$ removes more mass and angular momentum from the disk, reducing surface density and tangential velocity. Consequently, viscous dissipation decreases. Stronger outflows correlate with lower accretion rates (except at the outer edge, which is fixed), and slightly elevated radial velocities in the outer disk. The sound speed remains largely unaffected. Meanwhile, the Toomre $Q$ value increases with $\Lambda_0$, but remains below the GI threshold in our tested range. Thus, winds moderate disk build-up and can delay—but not entirely suppress—GI in the protodisk parameter space explored here.

The $M_{\rm disk}/M_{\rm env}$ ratio also declines with increasing $\Lambda_0$, indicating that outflows eject disk mass, some of which may be re-accreted by the envelope in realistic systems. Thus, outflows modulate disk mass accumulation efficiency.
As shown in Figures ~\ref{timeevol}–\ref{lambda}, $v_{\phi} \propto r^{-0.2}$, which matches the theoretical profile ($r^{-0.2}$) estimated for a thin disk model in Appendix B (see Eq.~\ref{b6}).

In Figure~\ref{gamma}, we test the thermodynamic sensitivity by varying $\gamma$. Raising $\gamma$ from 1 to 1.1 increases surface density, tangential velocity, and accretion rate, and by that, the total disk-to-mass envelope mass ratio. In contrast, the radial velocity, sound speed, and Toomre $Q$ all decrease. These effects reflect the stiffer pressure support in polytropic disks. Our results show that for $\gamma = 1$, the disk is gravitationally stable, while for $\gamma = 1.1$, it becomes unstable. Within our similarity framework, larger $\gamma$ tends to push $Q$ toward unity in the outer disk while leaving a broad annulus with $Q\!\lesssim\!1$, where GI provides the dominant transport.

To map out conditions where GI remains active, we construct a grid in $\alpha_{0}$–$\gamma$ space (Figure~\ref{grid}), computing the radially averaged Toomre parameter $\bar{Q}$. The models fall into three categories: (1) always stable ($\bar{Q} > 1$), (2) always unstable ($\bar{Q} < 1$), and (3) transitioning from stable to unstable. Our GI-based viscosity prescription is only valid in regimes where $\bar{Q} < 1$, delineating the parameter space in which our solutions are self-consistent. Accordingly, we interpret runs with $\bar{Q}\!\ge\!1$ as diagnostic rather than predictive for transport: they indicate where a different mechanism, e.g., magnetorotational instabilities (MRI), irradiation, or a point-mass potential would need to be included. Figure \ref{grid} shows that the models generally have $\bar{Q} < 1$ within $t = 1$ kyr.

% Figure~\ref{menv} illustrates the relationship between $M_{\rm disk}/M_{\rm env}$ and $M_{\rm env}$ across models with varying $\alpha_0$. As expected, lower $\alpha_0$ values—representing more massive and gravitationally unstable cores—lead to enhanced disk growth and higher disk-to-envelope mass ratios. \textbf{For context, we overlay observational points (Class~0 and I) reported by interferometric surveys; these are shown for orientation only and do not imply calibration of our protodisk solutions.} These studies report broad distributions in disk masses and radii, with Class 0 disks in particular exhibiting early onset of substantial disk formation—consistent with our model predictions.

% It is important to acknowledge that the disk masses reported in these studies may include small-scale structures that are not exclusively rotationally supported. In contrast, our modeled disks correspond to the protodisk stage and are assumed to be fully centrifugally supported. This difference in definition can systematically shift observationally inferred $M_{\rm disk}$ upward relative to our model values and should be kept in mind when reading Figures~\ref{alpha}–\ref{gamma}.

The overall result seen in Figures~\ref{timeevol} - \ref{lambda} is that there are notable power-laws in profiles of physical variables for models that remain nearly isothermal. Although there are small variations from model to model, we discuss the results in an average sense here. The $Q$ parameter remains nearly spatially uniform due to
the influence of the effective viscosity $\nu_{\rm GI}$,
%the $Q$-dependent viscosity, 
but there is a slightly increasing value versus radius, so we approximate $Q \propto r^{0.1}$ within the disk. The sound speed is slightly decreasing with radius and we adopt $c_{\rm s} \propto r^{-0.1}$. The surface density obeys approximately $\sigma \propto r^{-1.4}$. Inserting these scalings into Eq. (\ref{toomre}) and using Eq. (\ref{epicyclic}), we find that $\Omega \propto r^{-1.2}$, therefore $v_\phi \propto r^{-0.2}$. This scaling of $v_\phi$ is also consistent with centrifugal balance with a gravitational field $g_{\rm disk} \propto r^{-1.4}$ that emerges naturally from a surface density $\sigma \propto r^{-1.4}$. See Appendix \ref{app2} for details. 

Uncertainties in observationally inferred disk masses remain significant due to systematic dependencies on dust opacities, temperature assumptions, and simplified radiative transfer models. We revisit these limitations and their implications in Section~\ref{conclusion}. Overall, the parameter trends here identify a self-consistent window—set by $\alpha_0$, $\eta'$, $\Lambda_0$, and $\gamma$—in which GI and winds regulate the early protodisk before the epoch when the stellar potential and irradiation must be included explicitly.

% few different models, and plotting them for 1000 years.

\section{Angular Momentum Evolution}\label{angmom}

% \begin{figure}
% \begin{center}
% \includegraphics[width=\columnwidth,height=0.4\textheight]{new_figures/new_figure_8.pdf}
% \end{center}
% \caption{Time evolution of the specific angular momentum of the circumstellar disk, for different cloud parameters $\alpha_0$, up to a few kyr after the onset of protostar formation. Dashed lines correspond to solutions without any outflow.}
% \label{sangmom1}
% \end{figure}

\begin{figure*}[t]
\begin{center}
\includegraphics[width=0.95\textwidth]{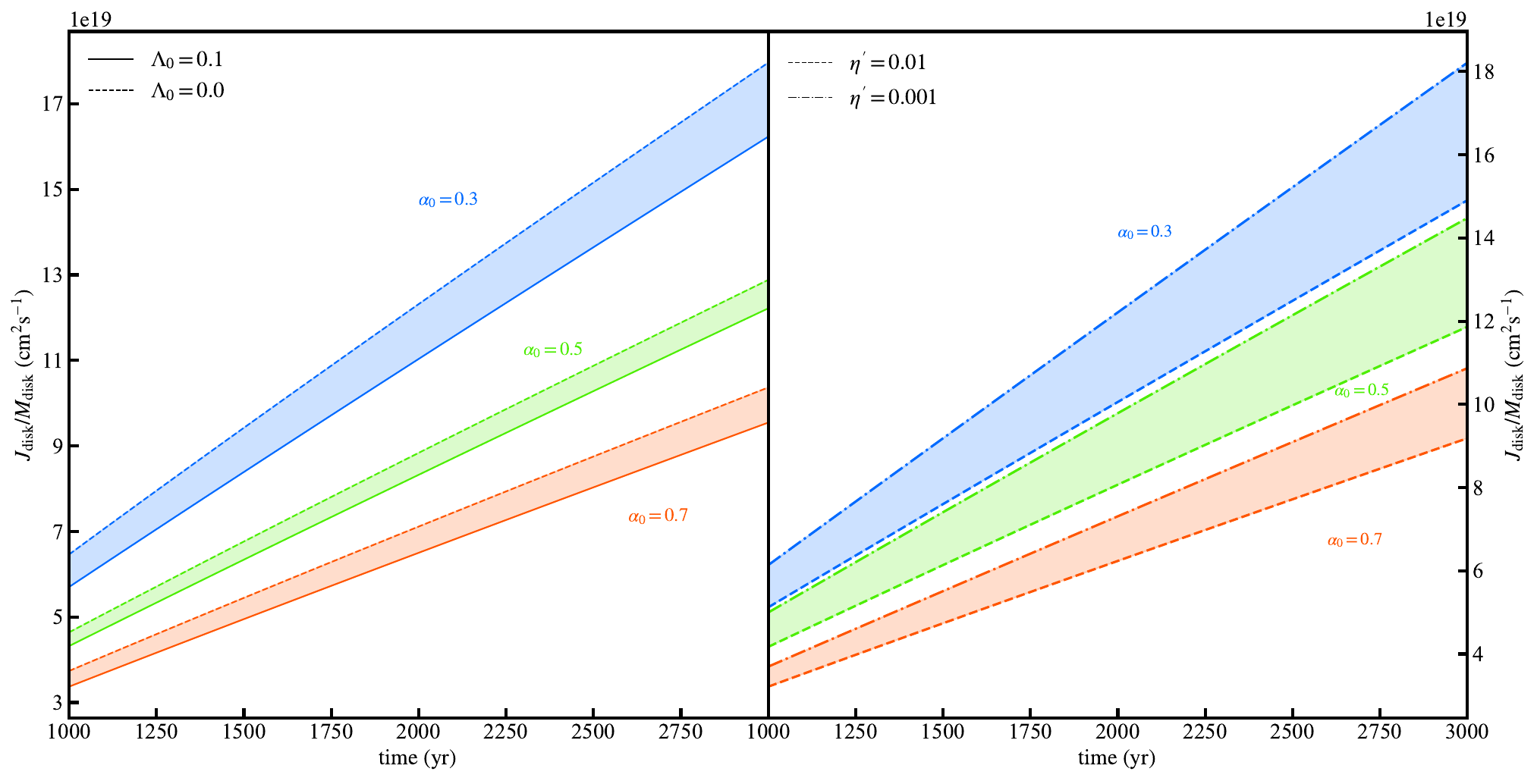}
\end{center}
\caption{Left: Time evolution of the specific angular momentum (units: $\rm 10^{19} cm^{2} s^{-1}$) of the circumstellar disk, for different  $\alpha_0$ and  $\Lambda_0$, up to 3 kyr after the onset of protostar formation. Dashed lines correspond to solutions without any outflow. Right: Time evolution of the specific angular momentum of the circumstellar disk, for different $\alpha_0$ and  $\eta^{\prime}$.}
\label{sangmom}
\end{figure*}

\begin{figure}
\begin{center}
\includegraphics[width=\columnwidth,height=0.5\textheight]{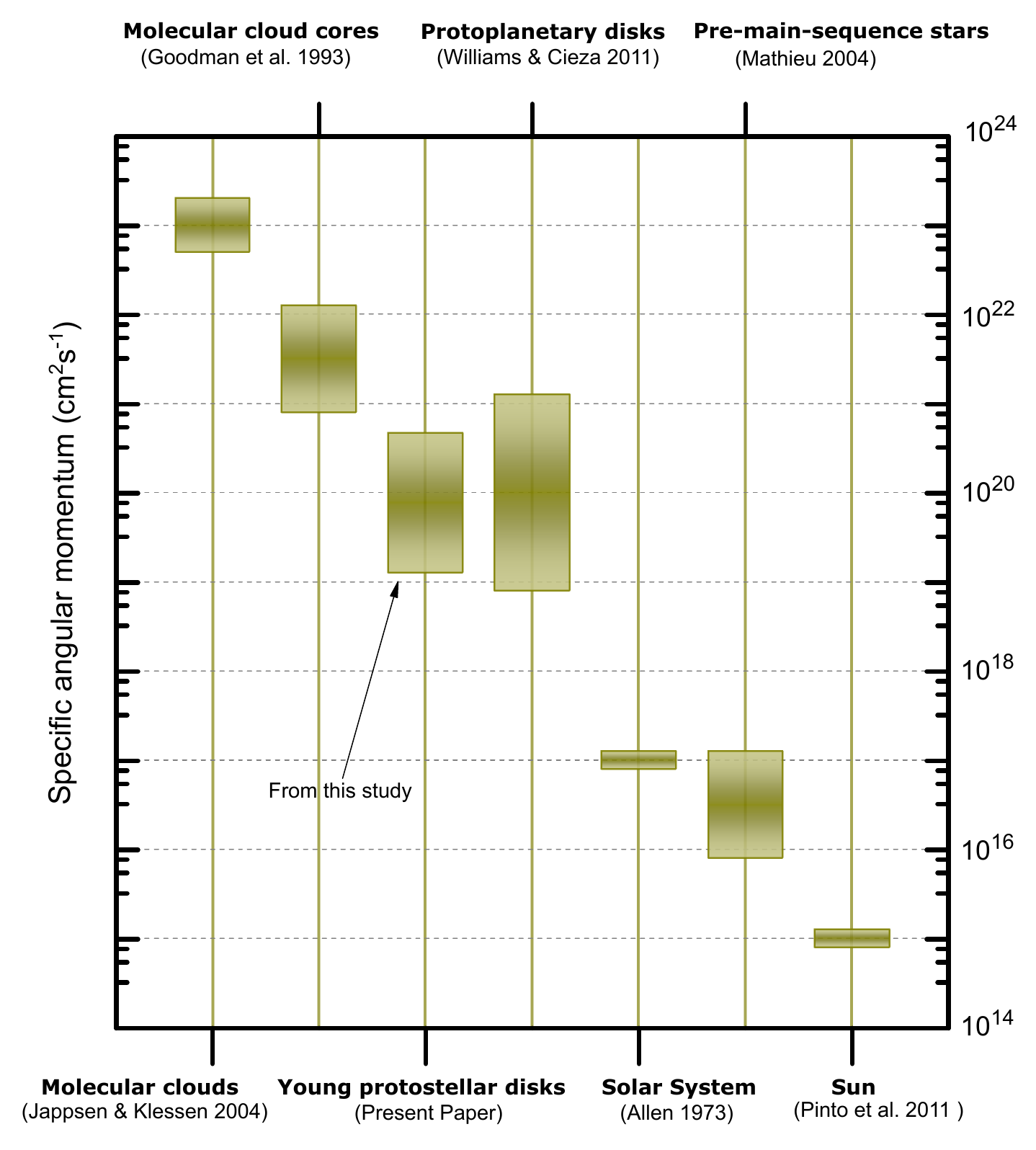}
\end{center}
\caption{Schematic representation of typical specific angular momenta over different scales and stages in stellar evolution, from molecular clouds \citep{goodman1993dense, jappsen2004protostellar}, to protoplanetary disks \citep{williams2011protoplanetary}, and finally to pre-main-sequence stars and the Sun \citep{allen1973astrophysical, mathieu2004rotation, pinto2011coupling}}
% include the refernces in the caption
\label{sangmom3}
\end{figure}

Angular momentum is transported from the faster inner disk to the slower outer regions via gravitational torques and disk winds. In the earliest phase, the angular momentum inherited from the collapsing core is transferred to the disk. We analyze the evolution of the specific angular momentum in the disk, where the total angular momentum is
\begin{equation}
J_{\rm disk}= \int r v_{\phi}(r) \sigma(r) ~ 2\pi r \, \mathrm{d}r\, ,
\label{eq51}
\end{equation}
and the mass-weighted specific angular momentum is $J_{\rm M}\!\equiv\!J_{\rm disk}/M_{\rm disk}$.

The left panel of Figure~\ref{sangmom} shows the time evolution of the specific angular momentum for different values of $\alpha_{0}$ and $\Lambda_0$ (outflows and no outflows cases), with other parameters fixed at their fiducial values. As the disk grows in mass and size, its specific angular momentum increases over time. This is the expected scaling in a self-gravitating disk, as both the enclosed disk mass and characteristic radius increase, the mass-weighted $J_{\rm M}$ rises because the outer disk stores the angular momentum transported from inner radii. This behavior is consistent with results from previous numerical simulations \citep{hayfield2011properties, joos2012protostellar} and reflects the buildup of angular momentum from sustained envelope accretion. 

% The order of magnitude of the specific angular momentum $J_{\rm M}$ in our models is $\sim 10^{19}$--$10^{20}$ cm$^{2}$ s$^{-1}$, consistent with observational estimates of Class 0 disks \citep{belloche2013observation} and theoretical models of protostellar evolution. For example, \citet{jappsen2004protostellar} estimate an average value of $\langle J_{\rm M} \rangle \approx 8 \times 10^{19}~{\rm cm^2~s^{-1}}$ for forming systems, which closely matches our results. Although \citet{hayfield2011properties} simulated a more massive ($50~{\rm M}_\odot$) parental cloud, the disk masses and angular momentum ranges produced in their models agree well with ours. \textbf{We use these comparisons as orientation only; our solutions are restricted to the star-free protodisk limit and are not calibrated to typical Class~0 systems.}

The presence of outflows leads to a measurable reduction in specific angular momentum over time, as shown by the difference between dashed and solid curves in the left panel of Figure~\ref{sangmom}. Physically, winds remove high–specific-angular-momentum material from the disk surface and carry an extra torque through the magnetic lever arm (parameter $l$ in our model), thereby reducing $J_{\rm M}$ even when mass loss is modest. This supports the interpretation that disk winds act as a secondary angular momentum loss channel, particularly relevant in the early phases \citep{lebreuilly2024influence}.

The right panel of Figure~\ref{sangmom} explores how $J_{\rm M}$ varies with $\eta^{\prime}$ and $\alpha_0$. Increasing $\eta^{\prime}$ reduces the specific angular momentum $J_{\rm M}$, as stronger gravitational torques enhance outward transport of the angular momentum. Lowering $\alpha_{0}$ increases the envelope mass-loading (Eq.~\ref{eq11}), which boosts $J_{\rm M}$ at fixed time by growing the disk and shifting more angular momentum to larger radii. These trends are again in line with theoretical expectations for GI-regulated disks \citep{lin1987grav, Kratter2010}.

Measurements on cloud scales typically yield specific angular momenta around $~ \sim 10^{21} - 10^{22}$ cm$^2$ s$^{-1}$ \citep{goodman1993dense}, greater than those of individual disks. Figure~\ref{sangmom3} provides a schematic of $J_{\rm M}$ values across spatial scales and evolutionary stages. Our models offer a self-consistent explanation for the several-orders-of-magnitude angular momentum reduction required during collapse and disk formation—a core challenge known as the ``angular momentum problem” in star formation theory \citep{spitzer2008physical}. In our picture, the protodisk absorbs the angular momentum of the core, GI torques redistribute it radially, and the winds remove a fraction, setting the initial conditions for later star-dominated evolution. The GI torques in our model lead to angular momentum redistribution and mass accretion in a smoothly varying manner that can be considered to be the time average of a process that will more realistically also include episodic bursts \citep{vorobyov2005origin,vorobyov2006burst,vorobyov2010}.

This behavior is also consistent with recent disk formation simulations \citep[e.g.,][]{tsukamoto2022fragmentation}
that show that the early circumstellar disk acts as a reservoir absorbing angular momentum from the collapsing core.
%redistributed during collapse. 
After the second collapse phase, the residual disk continues to accrete angular momentum and expands outward. We therefore interpret the increase of $J_{\rm M}$ with time as a robust feature of the protodisk stage; beyond this stage, once $M_\star\!\gtrsim\!M_{\rm disk}$ and irradiation becomes important, a different angular-momentum budget should be considered.

\section{Comparison with Magnetic and Hydrodynamic Models}\label{Simul}

\begin{figure}
\begin{center}
\includegraphics[width=1.0\columnwidth]{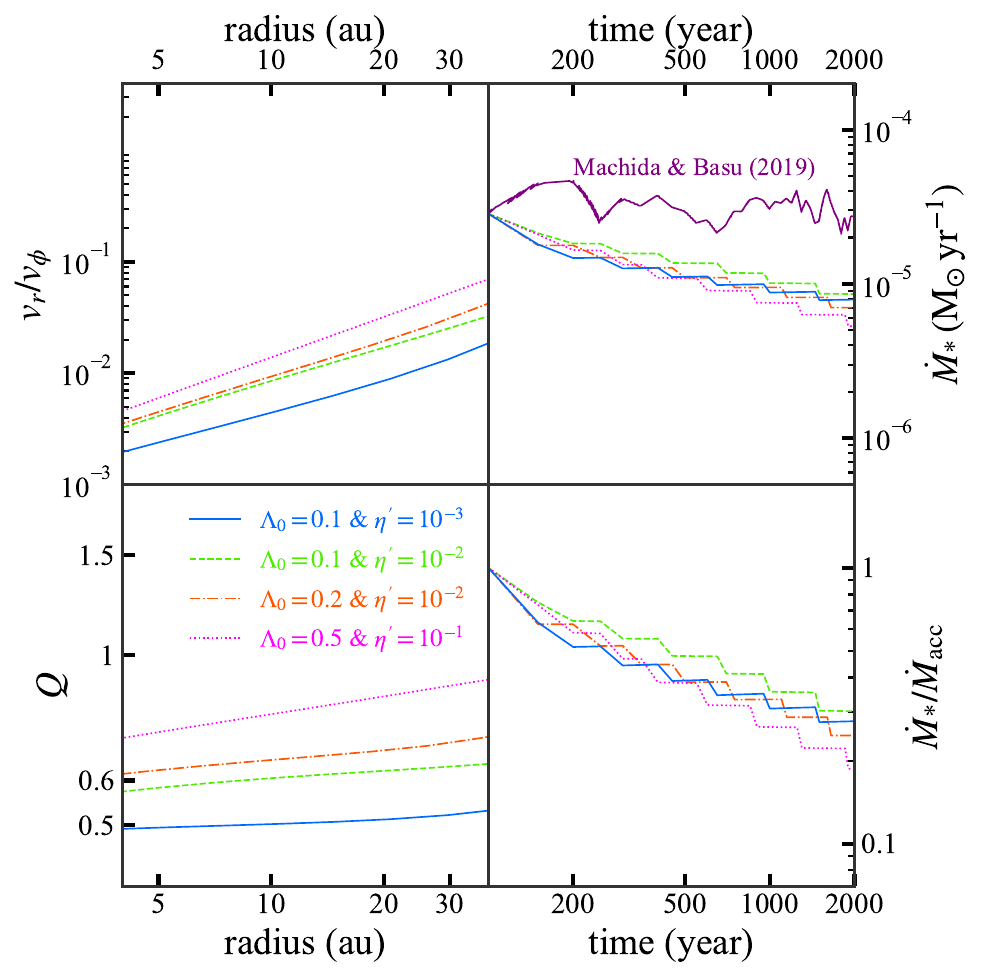}
\end{center}
\caption{Ratio of radial $v_r$ to tangential velocity $v_{\phi}$ along with the Toomre $Q$ as a function of radius at $t = 1000$ yr. The protostar mass accretion rate $\dot{M}_{*}$ and the ratio of protostar to disk accretion rates $\dot{M}_{*}/\dot{M}_{\rm acc}$ are plotted for the first 2000 yr of protostellar evolution. The mass accretion rate from the simulation of \cite{machida2019first} is shown for comparison.}
\label{Compact_figure}
\end{figure}

We explore the disk's temporal evolution until $t = 2000$ yr, by alternating both the outflow efficiency governed by the wind loading parameter $\Lambda_0$ and the angular momentum transport parameter $\eta^{\prime}$. In Figure~\ref{Compact_figure}, the left panels show physical quantities for a snapshot of the disk at $t = 1000$ yr and compare how different combinations of $\Lambda_0$ and $\eta^{\prime}$ affect the rotational motion, accretion, and stability of the disk. We assume the protostar accretion would dominate the inner accretion within ($r < 5$ au). Therefore, the right panels show 
%the time evolution of the protostellar core $(\dot{M_{\rm acc}} \,\, \, at\,\, r < 5 \, {\rm au})$ 
the protostar accretion rate $\dot{M}_{\rm *}$ (measured at $r = 5$ au) and the ratio of protostar accretion  ($\dot{M}_{\rm *}$) to protodisk accretion ($\dot{M}_{\rm acc}$).

The properties, including tangential velocity, disk mass, protostellar and protodisk accretion rate,  obtained by our self-similar model are in a similar range as \cite{machida2010formation}, involving purely
hydrodynamical simulations in the first $10^4$ yr after the protostar formation, when $M_{\rm disk} \gg M_*$. They differ slightly from \cite{machida2011effect} and \cite{machida2019first} due to the influence of magnetic fields that initiate magnetic braking and outflows. Because of the presence of magnetic fields in their models, the disk does not grow in size until the magnetic fields are diffused significantly. However, we observe that changing $\Lambda_0$ and $\eta^{\prime}$ leads to efficient mass outflow and angular momentum transport from the inner regions, bringing us closer to the early stage disk properties seen in MHD simulations.

Small values for $\eta^{\prime}$ and $\Lambda_0$ lead to ineffective angular momentum transport and weak outflows, respectively. This results in more mass and angular momentum retained within the inner regions of the disk.
%, allowing viscosity-induced gravitational instabilities to occur.
Therefore, the Toomre parameter $Q$ remains sufficiently within the unstable range even near the outer disk's boundary. This is recorded in the simulations of \cite{machida2010formation}. In their simulations, the absence of magnetic fields results in no magnetic braking and weak outflows. This leads to a steady increase in the disk's mass and size. The disk maintains average Toomre $Q$ lower than unity until the protostar grows in size and becomes comparable to the disk in mass.

In Figure~\ref{Compact_figure}, we plot physical parameters that have been extensively studied in early-stage protostellar-disk simulations \citep{machida2010formation,machida2011effect,machida2019first}. We plot the ratio of the radial velocity $v_r$ to the tangential velocity $v_\phi$ in the upper left panel. 
%The tangential velocity  $v_\phi$ is always greater than $v_r$. 
The ratio is always greater than unity. This is because of the slow accretion limit assumption in our model. The steady increase in the ratio of radial to tangential velocity is explained by the observations made in Figure~\ref{eta-prime} and Figure~\ref{lambda}. Increasing the GI viscosity parameter $\eta^{\prime}$ leads to higher accretion, and increasing the outflow parameter $\Lambda_0$ transports angular momentum of the protodisk to outer regions, causing the outer regions of the disk to have higher radial velocities.
Higher $\Lambda_0$ and $\eta^{\prime}$ lead to a complex interplay of physical effects. Higher $\Lambda_0$ continuously transfer mass and angular momentum out of the system through stronger winds, and the disk loses mass and becomes less prone to gravitational instabilities. However, higher $\eta^{\prime}$ induce gravitational instabilities and maintain efficient angular momentum transport, which causes gravitational torques to transfer mass into the inner regions of the disk. Because of the interplay between higher $\Lambda_0$ and $\eta^{\prime}$, no significant influence is seen in the inner accretion of the disk, resulting in a relatively unchanged protostar accretion rate.

\cite{machida2019first} simulated a collapsing core where magnetic fields slowly dissipate, causing angular momentum to be less effectively transferred and for outflows to decay. The disk surface density in their model gradually increases and causes GI and accompanying gravitational torques to operate. This is consistent with our model, where weaker outflows and higher surface density result in a disk more prone to GI. The protodisk mass remains greater than the protostellar mass, which is observed in the unmagnetized core collapse simulations performed by \citet{machida2010formation}.

In the lower left panel, we can see that increasing the outflow parameter and viscosity leads the system to greater stability and higher $Q$ values. This is because viscosity increases the angular momentum transport, which leads to the system losing mass. Since the induced GI viscosity depends on the disk mass, this will make the system more stable and lead to the Toomre parameter $Q$ being closer to stability. 

In the upper right panel, we compare our results of protostar accretion rates with that of \cite{machida2019first}, finding the values to be in a similar range. 
Although the model does not include a protostar, in the sense that the mass of the protostar is negligible ($M_{*} \ll M_{\rm disk}$), and does not contribute to the gravitational potential, the accretion from the disk to an inner region still happens.
In the lower right panel, the ratio of the protostar's mass accretion to the disk accretion is plotted. The accretion from the envelope to the disk is greater than the accretion from the disk to the protostar, and the ratio is always less than unity after a couple of hundred years of evolution, making it consistent with our model's assumptions. 

% Add a plot for the Mprotostar for differnt eta and lambda
\section{Conclusions}\label{conclusion}

% \begin{figure}
% \begin{center}
% \includegraphics[width=\columnwidth]{new_figures/disk_mass_vs_eta_and_lambda.jpg}
% \end{center}
% \caption{Schematic representation of typical specific angular momenta over different scales and stages in stellar evolution.}
% \label{M_disk_11}
% \end{figure}

% \begin{figure}
% \begin{center}
% \includegraphics[width=\columnwidth]{new_figures/number_density_fiducial_case.jpg}
% \end{center}
% \caption{Fiducial Case Number Density.}
% \label{M_disk_12}
% \end{figure}

We investigated the structure and early dynamical evolution of the immediate post--protostar-formation, self-gravity--dominated protodisk (i.e., $M_{\rm disk}\!\gg\!M_\star$). Our model neglects the gravitational force of the central protostar by restricting the analysis to the first few kyr of disk evolution, during which the disk mass dominates. In practice, we treat the stellar point-mass term and irradiation as dynamically minor and omit them from the equations; we therefore do not extrapolate our solutions to typical Class~0/I systems where $M_\star\!\gtrsim\!M_{\rm disk}$. We adopted a polytropic pressure-density relation and parameterized angular momentum loss via disk outflows. Gravitational instability (GI) was modeled through the effective viscosity prescription of \citet{lin1987grav}. A rough applicability window is the first $\lesssim 2\times10^{3}$~yr after protostar formation.

A central theme of this work is the link between the initial conditions of the molecular cloud core and the resulting disk properties. We introduced the parameter $\alpha_0$ to control the mass infall rate from the envelope to the disk and studied its effect alongside three other critical parameters: the polytropic index $\gamma$, the wind strength parameter $\Lambda_0$, and the efficiency of GI-induced angular momentum transport $\eta^{\prime}$. This mapping from $(\alpha_0,\gamma,\Lambda_0,\eta')$ to disk structure provides an analytic, self-similar framework for identifying where GI and winds regulate the protodisk.

%Although the formation of rotationally supported disks at early protostellar stages remains under debate, 
%A growing number of high-resolution observations now support their early presence \citep{gerin2017evidence, tobin2020class0}. Some studies (e.g., \citealt{enoch2009properties}) suggest that large disks form early in collapse, while others point to magnetic braking as a possible suppressor of disk growth \citep{li2011non}. In this study, we ignored magnetic fields, assuming that magnetic braking is weakened in dense, weakly ionized cores \citep{bergin2007cold, nakano2002mechanism}. This assumption is partially supported by chemical and microphysical arguments \citep{oppenheimer1974fractional} and may help explain why our model predicts higher $M_{\rm disk}/M_{\rm env}$ ratios than those inferred from some ALMA observations. Notably, \citet{zhao2016protostellar} and \citet{yen2016gas} suggest that the region of efficient magnetic braking may lie at radii $\gtrsim 100$~AU, possibly allowing disks to survive at smaller scales. \textbf{We emphasize that all observational comparisons here are for illustrative context only; the solutions themselves are intended for the protodisk stage.}

Our theoretical model predicts relatively high disk-to-envelope mass ratios, particularly for low $\alpha_0$ cases that represent more gravitationally unstable parental cores. At first glance, these values may appear higher than those inferred from dust continuum observations of Class~0 protostars. However, this discrepancy may be due to systematic biases in observational mass estimates.
\citet{tsukamoto2017impact} demonstrated through radiation hydrodynamic simulations that in gravitationally unstable disks, dust growth and depletion of small grains can significantly lower the millimeter-wave opacity. This leads to an underestimation of the disk mass by a factor of 3 to 5 when standard interstellar medium dust properties are assumed. Given that our regime targets GI-active, rapidly evolving protodisks, such opacity effects could naturally move observed $M_{\rm disk}$ estimates toward our predicted values.

Comparisons with early-stage protostellar-disk simulations provided insight into the applicability of our theoretical model. In our theoretical model the disk evolves with properties resembling hydrodynamical simulations in the first $10^3 $ to $10^4$ yr because we did not consider magnetic fields in the self-similar model. Adjusting the key parameters, specifically $\Lambda_0$ and $\eta^{\prime}$, induces efficient angular momentum transport to the outer-regions due to GI viscosity and stronger wind efficiencies, resulting in the protodisk having properties similar to those observed in MHD simulations.

Due to the inherent limitations of similarity solutions near boundaries, we did not model the detailed transition from disk to protostar. In our formulation, the disk mass includes the mass that will eventually accrete onto the central object. Likewise, fragmentation physics and radiative feedback are beyond the present scope and are deferred to future work.
Our theoretical analysis highlights several key trends:

\begin{itemize}
  \item Disks originating from more gravitationally unstable cores (lower $\alpha_0$) rotate more rapidly, exhibit higher accretion rates, and develop systematically higher surface densities. Such disks are more prone to GI, as indicated by lower Toomre $Q$ values.
  
  % \item Such disks are more prone to GI, as indicated by lower Toomre $Q$ values.
  % and they accumulate larger specific angular momenta over time.

  \item While gravitational torques and disk outflows act to extract angular momentum, continued envelope accretion replenishes it. This results in a net monotonic increase in specific angular momentum throughout the early disk evolution.
  % —consistent with previous studies \citep{hayfield2011properties, joos2012protostellar}.
  \item Increased efficiency of angular momentum transport (i.e., higher values of $\eta^{\prime}$) leads to more effective outward transfer of angular momentum, thereby reducing the amount retained 
  within the disk. 
  \item The centrifugally supported disk is characterized by a profile shallower than Keplerian in which we find the approximate relation $v_\phi \propto r^{-0.2}$. This follows from a surface density $\sigma$ and resulting gravitational field $g_{\mathrm{disk}}$ that both scale as $\propto r^{-1.4}$.
  %$ that scale as  where $\sigma$, $g_{\mathrm{disk}} \propto r^{-1.4}$ and $v_\phi \propto r^{-0.2}$, which could be shown by solving the gravitational field equation for a thin disk and comparing it to the spherical approximation}.
\end{itemize}

Together, these results underscore the importance of GI and envelope feeding in governing the dynamical evolution of protostellar disks. Outflows and internal GI torques lead to a significant loss of angular momentum that works toward resolving the angular momentum problem of star formation. Our model suggests that the structure, stability, and mass content of the protodisk are primarily determined by the thermodynamic state of the parental core and the efficiency of GI and winds; beyond this stage, as $M_\star$ grows and the irradiation strengthens, a different (star-dominated) framework must be applied.

\begin{acknowledgments}
S. B. is supported by a Discovery Grant from the Natural Sciences and Engineering Council of Canada.
\end{acknowledgments}

% \acknowledgements
% E. N. gratefully acknowledges the hospitality of the School of Astronomy (SoA), IPM, Iran, where part of this work was completed. S. D. is supported by a Marie-Curie Intra European Fellowship under the European Community's Seventh Framework Program FP7/2007-2013 grant agreement no 627008.

\bibliography{disk_apjNotes_update}{}
\bibliographystyle{aasjournal} 

\appendix

\section{Similarity Transformation} \label{app1}

The following dimensionless similarity variables describe the time-dependent evolution of the physical quantities in our model.

Surface density:
\begin{equation}
\sigma(r,t) = (2\pi)^{-1} K^{1/2} G^{-(1+\gamma)/2} t^{-\gamma} \Sigma(x).
\end{equation}

Volume density:
\begin{equation}
\rho(r,t) = (4\pi\gamma)^{-1/\gamma} G^{-1} t^{-2} \Sigma^{2/\gamma}(x).
\end{equation}

Pressure:
\begin{equation}
p(r,t) = (4\pi\gamma)^{-1} K G^{-\gamma} t^{-2\gamma} \Sigma^2(x).
\end{equation}

Sound speed:
\begin{equation}
c_{\rm s}(r,t) = (4\pi)^{(1-\gamma)/(2\gamma)} \gamma^{1/(2\gamma)} K^{1/2} G^{(1-\gamma)/2} t^{1-\gamma} \Sigma^{(\gamma-1)/\gamma}(x).
\end{equation}

Radial velocity:
\begin{equation}
v_{r}(r,t) = K^{1/2} G^{(1-\gamma)/2} t^{1-\gamma} V_{r}(x).
\end{equation}

Tangential velocity:
\begin{equation}
v_{\phi}(r,t) = K^{1/2} G^{(1-\gamma)/2} t^{1-\gamma} V_{\phi}(x).
\end{equation}

Vertical velocity from outflows:
\begin{equation}
v_{z}^{+}(r,t) = K^{1/2} G^{(1-\gamma)/2} t^{1-\gamma} V_{z}^{+}(x).
\end{equation}

Specific angular momentum:
\begin{equation}
j(r,t) = K G^{1-\gamma} t^{3-2\gamma} J(x).
\end{equation}

Vertical scale height:
\begin{equation}
H(r,t) = (4\pi)^{(1-\gamma)/\gamma} \gamma^{1/\gamma} K^{1/2} G^{(1-\gamma)/2} t^{2-\gamma} \Sigma^{(\gamma-2)/\gamma}(x).
\end{equation}

Effective viscosity:
\begin{equation}
\nu(r,t) = K G^{1-\gamma} t^{3-2\gamma} \nu^{\prime}(x).
\end{equation}

Cumulative mass interior to radius $r$:
\begin{equation}
M_{r}(r,t) = K^{3/2} G^{(1-3\gamma)/2} t^{4-3\gamma} \mathcal{M}_{x}(x).
\label{eq:Mrp}
\end{equation}

Mass accretion rate:
\begin{equation}
\dot{M}_{\rm acc}(r,t) = -2\pi r \sigma v_{r} = K^{3/2} G^{(1-3\gamma)/2} t^{3(1-\gamma)} \dot{\mathcal{M}}_{\rm acc}(x).
\end{equation}

Mass loss rate due to outflows:
\begin{equation}
\dot{M}_{\rm w}(r,t) = K^{3/2} G^{(1-3\gamma)/2} t^{3(1-\gamma)} \dot{\mathcal{M}}_{\rm w}(x).
\end{equation}

Mass loss per unit area (from Eq.~\ref{eq5}):
\begin{equation}
\dot{\sigma}_{\rm w}(r,t) = (4\pi)^{-1} K^{1/2} G^{-(1+\gamma)/2} t^{-\gamma-1} \Sigma(x) \Gamma(x)
\label{eq:mwp}.
\end{equation}

Auxiliary dimensionless functions:
\begin{equation}
J(x) = x V_{\phi}(x),
\end{equation}

\begin{equation}
\mathcal{M}_{\rm x}(x) = \int_0^x \Sigma(x')\, x' dx'
\label{eq:Mx},
\end{equation}

\begin{equation}
\dot{\mathcal{M}}_{\rm w}(x) = \int_0^x \Sigma(x') \Gamma(x') x' dx' ,
\end{equation}

\begin{equation}
\dot{\mathcal{M}}_{\rm acc}(x) = \frac{\Sigma\, x\, v_{r}}{3\gamma - 4},
\end{equation}

\begin{equation}
\Gamma(x) = (4\pi)^{(\gamma - 1)/\gamma} \gamma^{-1/\gamma} \Sigma^{2(1-\gamma)/\gamma}(x) \,\Lambda(x), \quad \Lambda(x) = \Sigma\, V_{z}^{+}(x)
\label{a19}.
\end{equation}

\section{Disk Gravitational Field}
\label{app2}

%\citet{morton1994} and
\citet{basu1997} estimated that the gravitational collapse of a magnetized cloud core with supercritical mass-to-flux ratio leads to a surface density profile:
%supercritical core collapse, the surface density of the gas the profile:

\begin{equation}
\sigma(r) = \frac{\sigma_0}{\sqrt{1 + (r/R)^2}} \, ,
\label{b1}
\end{equation}
where $\sigma_0$ is the central surface density and $R$ is the instantaneous size of a central plateau of surface density. The asymptotic ($r \gg R$) form of this profile is $\propto r^{-1}$.

 According to \citet{toomre1963}, the gravitational field of a thin disk with surface density profile $\sigma(r)$ is
\begin{equation}
g_{\rm disk}(r) = -2\pi G \int_0^\infty dr'r'\sigma(r')\,M(r,r'),
\label{b2}
\end{equation}
where
\begin{eqnarray}
M(r, r') & = & \frac{2}{\pi} \frac{d}{dr}\left[\frac{1}{r_>} K\left(\frac{r_<}{r_>}\right) \right],\\
r_< & = & \min \, (r, r'), \nonumber\\
r_> & = & \max \,(r,r'),\nonumber
\label{b3}
\end{eqnarray}
and $K$ is the complete elliptic integral of the first kind. Integrating equation (\ref{b2}) yields the gravitational field that corresponds to the surface density of the protostellar disk. For the profile given by Eq. (\ref{b1}) the gravitational field is \\
\begin{equation}
g_{\rm disk}(r) = \frac{-2 \pi G \, \sigma_0 \, r \,} {\, (r^2 + R^2)^{1/2} \left[ 1 + \sqrt{1 + \left( \frac{r}{R} \right)^2}\right]}\, ,
\end{equation}
as in \cite{basu1997}. The asymptotic form $(r \gg R)$ of this profile is 
\begin{equation}
   g_{\rm disk}(r) = -  \frac{2 \pi G \, \sigma_0 \, R \,}{r} \, .
   \label{gdiskasymp}
\end{equation}
Integrating Eq. (\ref{b1}) yields an enclosed mass within a radius $r$ equal to 
\begin{equation}
    M(r) = 2 \pi \sigma_0 R^2 \left[ \sqrt{1 + \left( \frac{r}{R} \right)^2} - 1\right] \, ,
\end{equation}
which has asymptotic ($r \gg R$) form
\begin{equation}
    M(r) =  2 \pi \sigma_0 R \, r \, .
    \label{Masymp}
\end{equation}
Comparing Eq. (\ref{gdiskasymp}) and Eq. (\ref{Masymp}), the asymptotic gravitational field can be written as 
\begin{equation}
    g_{\rm disk}(r) = - \frac{GM(r)}{r^2} \, .
    \label{gsphere}
\end{equation}
This partially justifies our model assumption in Eq. (\ref{eq2}) that the gravitational field of the thin disk can be written in the same way as for a spherical mass distribution. However, a disk generally does not have a gravitational field that behaves in the same way as that of spherical object. The net gravitational field is determined by matter from both interior and exterior to a radius $r$. Power-law profiles $\sigma \propto r^{-p}$ with $p<2$ yield gravitational fields that can be approximated by Eq. (\ref{gsphere}), and there is an exact match when $p=1$ as shown above. In our models presented in this paper we generally find $p \simeq 1.4$. Hence we explore this case numerically below.\\
%The surface density profile derived in \citet{basu1994} has the asymptotic solution profile varying with radius as $r^{-3}$. Therefore, depending on the characteristics of the protostellar disk, its surface density has the following profile:

A surface density profile generalized from Eq. (\ref{b1}) is
\begin{equation}
\sigma(r) = \frac{\sigma_0}{\sqrt{1 + (r/R)^n}}\, ,
\label{b4}
\end{equation}
where $n$ can have different values. We choose $n=2.8$ so that the asymptotic profile $\propto r^{-1.4}$.
%where $n = 2,3,4 \,\,...$ is decided based on the different characteristics of the disk.\\
%As seen in Figures~\ref{timeevol} -\ref{gamma}, the self-similar solution has a surface density $\sigma \propto r^{-1.5}$. Therefore, for this model focused on the \textit{protodisk} stage, the surface density profile is approximated by Eq. (\ref{b4}) with $n=3$:
%\begin{equation}
%\sigma(r) = \frac{\sigma_0}{\sqrt{1 + (r/R)^3}}.
%\label{b8}
%\end{equation}
%Asymptotically $(r \gg R)$ Eq.~(\ref{b8}) has a profile that varies as $r^{-1.5}$ as observed in the self-similar solutions. 
We solve for the gravitational field using Eq. (\ref{b2}) by applying the methods described in \cite{morton1994} and \cite{basu1994}.
%Solving the above equation through the in-house numerical solver by \citet{basu1994} gives us a relation where the gravitational potential of the disk $g_{disk}$ has the relation:\\
The asymptotic form of the calculated gravitational field is
\begin{equation}
    g_{\rm disk}(r) \propto r^{-1.4} ,
    \label{b9}
\end{equation}
%The centripetal acceleration at a radius $r$ relates to the gravitational field of the disk and the protostar as:
%\begin{equation}
%\frac{v_{\phi}^2}{r} = -g_* - g_{\rm disk},
%\label{b4}
%\end{equation}
%where $g_*$ is the gravitational field of the protostar and $g_{\rm disk}$ is the gravitational field of the disk.\\
%In our self-similar analytical model, we assumed that the mass of the protostar is much less than the disk mass ($M_{\rm disk} \gg M_{\rm protostar}$) throughout the simulation cycle. Therefore, the disk's gravitational potential should be much greater than the star's gravitational potential for larger radii and during the entire span of the very early-stage evolution ($g_{\rm disk} \gg g_{*}$). Therefore, the equation simplifies to
Using the numerical solution for $g_{\rm disk}$ in the relation
\begin{equation}
\frac{v_{\phi}^2}{r} = - g_{\rm disk}\, ,
\label{b11}
\end{equation}
we find that
%\begin{equation}
%\frac{v_{\phi}^2}{r} \propto r^{-1.5},
%\end{equation}\\
%and after simplification, the profile for tangential velocity $v_{\phi}$ with $r$ radius is: \\
\begin{equation}
    v_{\phi} \propto r^{-0.2}.
\label{b6}
\end{equation}

\renewcommand{\thefigure}
{B\arabic{figure}}
\setcounter{figure}{0}
\begin{figure}
\begin{center}
\includegraphics[width=0.7\columnwidth]{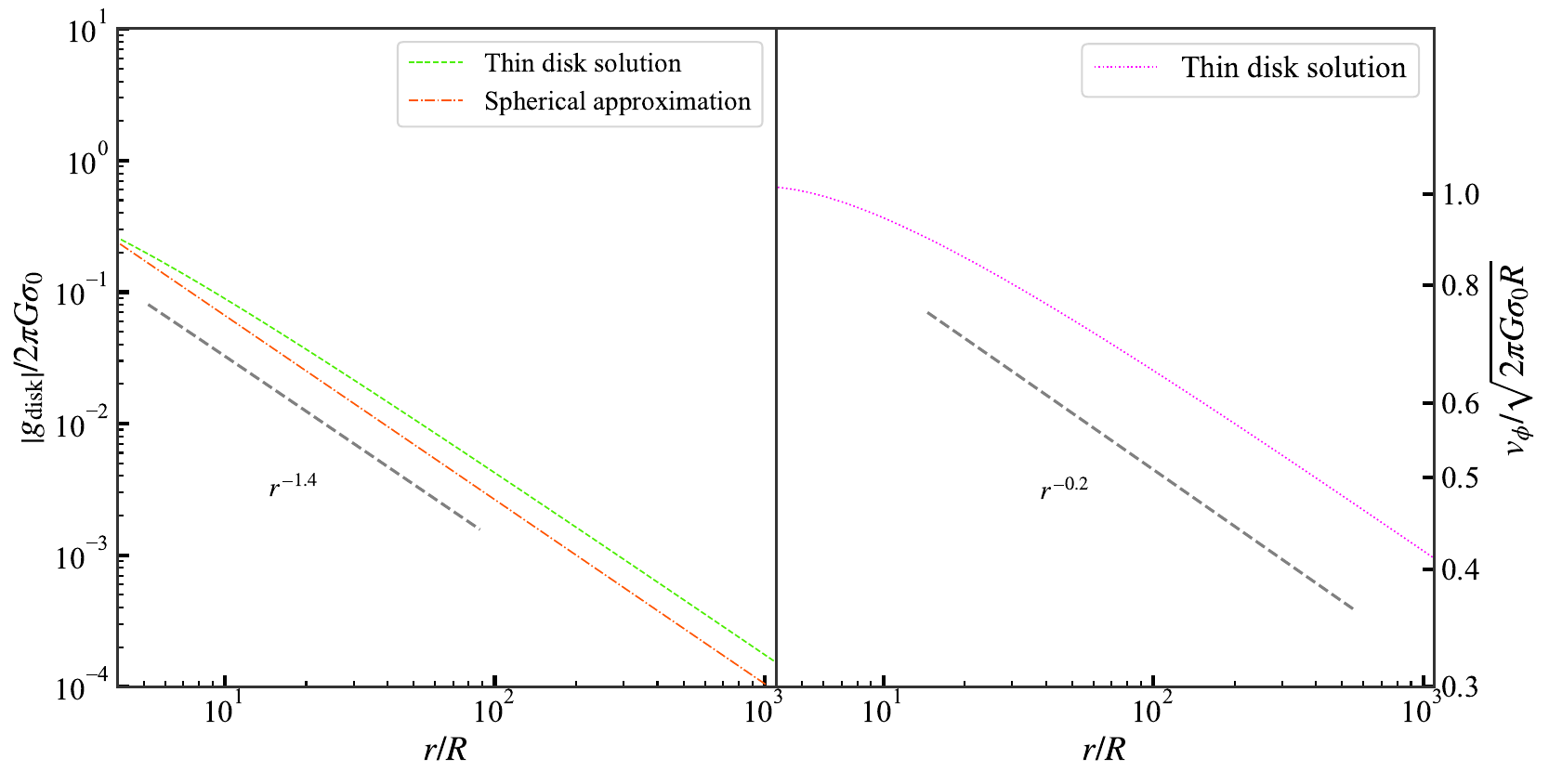}
\end{center}
\caption{Left: The radial profiles of the normalized gravitational field $g_{\rm disk}$ for the numerical solution for a thin disk using Eq. (\ref{b2}) and by using the spherical approximation $g_{\rm disk} = - GM(r)/r^2$. Right: Normalized tangential velocity $v_{\phi}$ radial profile calculated from our numerical solver for the thin-disk model (see Eq.~\ref{b2}). Regions at larger radii follow a radial profile $\propto r^{-0.2}$, which is in excellent agreement with the self-similar model presented in this study. The surface density $\sigma$ follows Eq. (\ref{b4}) with $n=2.8$.}
% include the refernces in the caption
\label{Appendix_plot}
\end{figure}

% \begin{figure}
% \begin{center}
% \includegraphics[width=0.35\columnwidth,height=0.35\textwidth]{new_figures/vphi_radial_profile.pdf}
% \end{center}
% \caption{The radial profile of normalized tangential velocity $v_{\phi}$ calculated from our numerical solver for the thin-disk model (see Eq.~\ref{b2}). Regions at larger radii follow a radial profile $\propto r^{-0.2}$, which is in excellent agreement with the self-similar model presented in this study.}
% % include the refernces in the caption
% \label{vphi_plot}
% \end{figure}
Figure~\ref{Appendix_plot} showcases the radial profiles of the normalized gravitational field and tangential velocity numerically calculated for the thin-disk model with a surface density having an asymptotic radial profile $r^{-1.4}$. Both the numerically calculated profiles match the radial profiles extracted from the solutions of the self-similar model (Figures~\ref{timeevol}-~\ref{lambda}). These profiles are shallower than the Keplerian radial profiles observed with systems dominated by a central stellar mass. We also compare the gravitational field using a spherical approximation (assumed in Eq.~\ref{eq2}) with the thin disk solution. The two differ on average by a small factor of $\simeq$ 1.5, with the thin-disk field being stronger since all of the mass is in the plane of the disk.

\clearpage

\end{document}